\PassOptionsToPackage{table,xcdraw}{xcolor}
\documentclass[12pt,a4paper]{article}
\pdfoutput=1

\usepackage{geometry}
\usepackage{comment}
\usepackage[numbers,sort&compress]{natbib}
\usepackage{amsmath}
\usepackage{amssymb}
\usepackage[dvips]{graphicx}
\usepackage{pstricks}
\usepackage{bm}
\usepackage{pbox}
\usepackage{placeins}
\usepackage{graphicx}
\usepackage{caption}
\usepackage{subcaption}
\usepackage[T1]{fontenc}
\usepackage{footnote}
\usepackage{pdfpages}
\usepackage{hhline}
\usepackage{multirow}
\usepackage{enumitem}
\usepackage{slashed}
\usepackage[nottoc,notlot,notlof]{tocbibind}
\usepackage[title,titletoc]{appendix}
\usepackage{dsfont}
\usepackage{pifont}
\allowdisplaybreaks
\usepackage{cases}
\usepackage{tikz}
\usepackage[symbol]{footmisc}
\usepackage{physics}
\usepackage{soul}

\usepackage{booktabs}
\usepackage{longtable}

\renewcommand{\thetable}{\Roman{table}}

\def\equationautorefname~#1\null{Eq.\,(#1)\null}

\definecolor{MyDarkBlue}{rgb}{0.1, 0.1, 0.8}
\definecolor{MyLightBlue}{rgb}{0.22,0.51,0.9}
\definecolor{MyGreen}{rgb}{0.0, 0.5, 0.0}
\definecolor{BrickRed}{rgb}{0.8, 0.25, 0.33}
\usepackage[colorlinks=true,linkcolor=blue,citecolor=MyDarkBlue,
urlcolor=MyLightBlue,bookmarksnumbered=true,bookmarksopen]{hyperref}
\hypersetup{colorlinks, citecolor=blue,linkcolor=blue, urlcolor=MyLightBlue}

\begin{document}
\vspace*{-0.2in}
\begin{flushright}

\end{flushright}
\vspace{0.5cm}
\begin{center}
{\Large\bf   
Non-Renormalizable SU(5) GUTs: \\ Leptoquark-Induced Neutrino Masses
}
\end{center}

\vspace{0.5cm}
\renewcommand{\thefootnote}{\fnsymbol{footnote}}
\begin{center}
{\large
{}~\textbf{Ilja Dor\v{s}ner$^{1,2}$}\footnote[1]{ E-mail: \textcolor{MyLightBlue}{dorsner@fesb.hr}}, 
{}~\textbf{Mijo Matkovi\'c$^{1}$}\footnote[3]{ E-mail: \textcolor{MyLightBlue}{mijo.matkovic@fesb.hr}}, 
{}~\textbf{Shaikh Saad$^2$}\footnote[2]{E-mail:  
\textcolor{MyLightBlue}{shaikh.saad@ijs.si}
}
}
\vspace{0.5cm}

{\em $^1$ University of Split, Faculty of Electrical Engineering, Mechanical Engineering and\\ Naval Architecture in Split, Ru\dj era Bo\v{s}kovi\'{c}a 32, HR-21000 Split, Croatia}\\
{\em $^2$Jožef  Stefan Institute, Jamova 39, P.\ O.\ Box 3000, SI-1001
  Ljubljana, Slovenia}
\end{center}

\renewcommand{\thefootnote}{\arabic{footnote}}
\setcounter{footnote}{0}
\thispagestyle{empty}

\begin{abstract}
We revisit the doublet–triplet splitting problem within the $SU(5)$ gauge group framework to advocate a viable regime with the light scalar leptoquark of the doublet–triplet splitting notoriety that is compatible with the current experimental bounds on partial proton decay lifetimes. We explicitly demonstrate, through a consistent use of higher-dimensional operators, how to implement suppression of baryon number violating interactions of the aforementioned color triplet. Our study thus offers an alternative approach to the doublet-triplet splitting problem as it removes a need for an extreme mass hierarchy between the partners residing in the same representation. We furthermore pursue two different extensions of two distinct symmetry breaking scenarios of $SU(5)$, one with a $24$-dimensional representation and the other one with a $75$-dimensional representation, to produce comparative study of novel consequences for the gauge coupling unification and the one-loop level neutrino mass generation. Our results point towards qualitatively novel $SU(5)$ scenarios, where the light scalar leptoquarks, responsible for the neutrino mass generation, might be even accessible at  colliders and thus serve as an accelerator accessible portal to the high-scale physics.
\end{abstract}
\newpage
{\hypersetup{linkcolor=black}
\tableofcontents}
\setcounter{footnote}{0}

\section{Introduction}

In the Georgi–Glashow~\cite{Georgi:1974sy} grand unified theory (GUT) proposal, the Standard Model (SM) Higgs doublet and a color triplet scalar leptoquark arise from the same $5$-dimensional representation of $SU(5)$. While the Higgs doublet is responsible for the electroweak symmetry breaking as well as the mass generation of the SM fields, its color triplet partner can mediate proton decay, thus posing a significant model building challenge. This issue is actually the source of the so-called doublet–triplet splitting problem~\cite{Dimopoulos:1981zb,Sakai:1981gr}. Namely, since the triplet mass should be near the gauge coupling unification scale, while the Higgs doublet must be at the electroweak scale, the generation of  such a tremendous mass hierarchy is considered to be one of the most persistent difficulties in construction of realistic theories.

In this work, we revisit an approach~\cite{Dorsner:2024seb} that challenges a need for such an extreme mass splitting. Namely, we explore a framework in which the color triplet can remain light --- possibly even within the reach of current and/or future colliders --- while still being completely insensitive to experimental bounds on partial proton decay lifetimes. We demonstrate, through the introduction of higher-dimensional operators, how the dangerous couplings that are responsible for baryon number violation can be suppressed with ease. (For alternative approaches to the proton decay suppression of interest, see Refs.~\cite{Nandi:1982ew,Berezinsky:1983va,Dvali:1992hc,Dvali:1995hp,Berezhiani:1998hg,Bajc:2002bv,Bajc:2002pg,FileviezPerez:2004hn,Dorsner:2004jj,Dorsner:2004xa,Dorsner:2005ii,Dorsner:2005fq,Nath:2006ut,Dorsner:2012nq, Dorsner:2012uz,Fornal:2017xcj}.) 

It needs to be stressed that we are not solving the doublet-triplet splitting problem as our proposal calls for substantial cancellations between dimensionless parameters in the Yukawa sector of the theory. However, our approach can yield collider-accessible triplet that directly couples to the SM fermions.
This clearly opens a door to rich low-energy phenomenology as light color triplet scalars should have observable consequences at current or upcoming experiments. Our results thus offer a shift in perspective. Namely, the color triplet may be a viable and testable component of a predictive GUT scenario instead of a theoretical nuisance to be decoupled at all cost. Moreover, the same scalar leptoquark can play a pivotal role in generating neutrino masses through the one-loop level quantum corrections as we discuss in detail later on. 

This work extends the original proposal~\cite{Dorsner:2024seb} by exploring new directions that enhance both its theoretical thoroughness and phenomenological richness. More specifically, we examine an alternative realization of the $SU(5)$ symmetry breaking mechanism using a $75$-dimensional scalar representation~\cite{Hubsch:1984pg} instead of the more conventional $24$-dimensional one. This substitution leads to a qualitatively different symmetry breaking pattern and has far-reaching implications for the structure of the theory. We investigate the resulting changes with regard to the scalar spectrum, gauge coupling unification, and proton stability. 
We accordingly provide a critical comparison between the $24$-dimensional and $75$-dimensional symmetry breaking scenarios, with particular attention to their differing impact on proton stability, doublet-triplet splitting, and the structure of higher-dimensional terms. We, furthermore, identify the simplest $SU(5)$ scenarios for the neutrino mass generation at the one-loop level, where the scalar leptoquark, an $SU(5)$ partner of the Higgs doublet, might reside at the scale accessible at colliders.  

Our work is organized as follows. In Sec.\ \ref{sec:24} we revisit the light color triplet regime when the $SU(5)$ gauge group is broken down to $SU(3) \times U(1)_\mathrm{em}$ with a $24$-dimensional representation and a $5$-dimensional scalar representation. Sec.\ \ref{sec:75} contains an analysis of the symmetry breaking scenario with a $75$-dimensional representation and a $5$-dimensional scalar representation. The neutrino mass generation, at the one-loop level, is discussed at length in Sec.\ \ref{sec:neutrinos} while Sec.\ \ref{sec:colliders} addresses potential experimental signatures of one of the scenarios under consideration. We briefly conclude in Sec.\ \ref{sec:conclusions}.

\section{24-Higgs}
\label{sec:24}

We first address the Georgi-Glashow scenario, where the $SU(5)$ gauge symmetry is broken down to the $SU(3) \times SU(2) \times U(1)$ gauge group with a $24$-dimensional scalar representation. The results we present in what follows dovetail with and substantially expand on the material presented in our previous study~\cite{Dorsner:2024seb}.

\subsection{Yukawa couplings}
The most relevant input for our discussion of  potential decoupling of the color triplet from proton decay signatures are the exact structure of the vacuum expectation values (VEVs) of $24$-dimensional and $5$-dimensional scalar representations and the associated $SU(5)$-invariant contractions in the Yukawa sector of the theory. Recall, the decomposition of $5_H$ under the SM gauge group is
\begin{align}    5_H&=T(3,1,-1/3)+D(1,2,1/2),
\end{align}
where $T$ is the color triplet scalar and $D$ is the Higgs boson doublet of the SM. Consequentially, $24_H$ decomposes under the SM gauge group as
\begin{align}    24_H&=\Phi_1(1,1,0)+\Phi_2(1,3,0)+\Phi_3(8,1,0)+
   \Phi_4(3,2,-5/6)+\Phi_4^*(\overline{3},2,5/6).
\end{align}
Note, to uniquely denote an $SU(5)$ representation we will use its dimensionality and additionally introduce subscripts $H$ or $F$ to specify whether a given representation contains scalars or fermions. 

The VEVs that sequentially accomplish the breaking of $SU(5)$ gauge group down to $SU(3) \times U(1)_\mathrm{em}$ are 
\begin{align}
\langle 24_H \rangle & =v_{24} \mathrm{diag}\left( -1, -1, -1, 3/2, 3/2 \right), \label{eq:24vev}
\\ \langle 5_H \rangle & = (0 \quad 0 \quad 0 \quad 0 \quad v_{5}/\sqrt{2})^T \label{eq:5vev},
\end{align}
while the interaction lagrangian reads
\begin{align}
&\mathcal{L}_Y= 
10_F^{\alpha ij}\bigg\{
Y_{d\alpha\beta}  \overline{5}_{Fi}^\beta 5^*_{Hj} +
\frac{1}{\Lambda} Y_{1\alpha\beta}  \overline{5}_{Fi}^\beta 5^*_{Hk} 24_{Hj}^k
+
\frac{1}{\Lambda} Y_{2\alpha\beta}  \overline{5}_{Fk}^\beta 5^*_{Hi}   24_{Hj}^k
\bigg\}
\nonumber\\&+
10_F^{\alpha ij}10_F^{\beta kl} 5_H^m
\bigg\{
Y_{u\alpha\beta}   \epsilon_{ijklm}
+
\frac{1}{\Lambda} Y_{3\alpha\beta}   24_{Hm}^n \epsilon_{ijkln}
+
\frac{1}{\Lambda} Y_{4\alpha\beta}  24_{Hk}^n \epsilon_{ijlmn}
\bigg\}+\mathrm{h.c.},
\label{eq:lagrangian_24}
\end{align}
where parameter $\Lambda$ represents a cutoff scale of the theory. The lagrangian of Eq.\ \eqref{eq:lagrangian_24} comprises all possible contractions between the SM fermions in $10_F^\alpha$ and fermions in either $10_F^\beta$ or $\overline{5}_{F}^\beta$ that are of dimensions $d=4$ and $d=5$. Here, $Y_d$, $Y_1$, $Y_2$, $Y_u$, $Y_3$, and $Y_4$ are Yukawa coupling matrices with complex entries, $\alpha, \beta =1,2,3$ are flavor indices while $i,j,k,l,m=1,\ldots,5$ are $SU(5)$ indices.

\subsection{Mass matrices}

The mass matrices of the SM fermions that populate $10_F^\beta$ and $\overline{5}_{F}^\beta$, as given by Eqs.\ \eqref{eq:24vev}, \eqref{eq:5vev}, and \eqref{eq:lagrangian_24}, are~\cite{Dorsner:2024seb}
\begin{align}
M_E &=v_5 \bigg\{\frac{1}{2}Y_d +\frac{3}{4} Y_1 \epsilon_{24}  - \frac{3}{4} Y_2 \epsilon_{24} \bigg\},\label{eq:uncorrected_E_24}\\
M_D &=v_5 \bigg\{\frac{1}{2}Y_d^T +\frac{3}{4} Y_1^T \epsilon_{24} + \frac{1}{2} Y_2^T \epsilon_{24} \bigg\}, \label{eq:uncorrected_D_24}\\
M_U &=v_5 \bigg\{  \sqrt{2}\left(Y_u+Y_u^T\right) + \frac{3}{\sqrt{2}}\left(Y_3+Y_3^T\right) \epsilon_{24}
 +  \left(\frac{1}{2\sqrt{2}} Y_4 -\sqrt{2}Y_4^T\right)\epsilon_{24}  \bigg\},\label{eq:uncorrected_U_24}
\end{align}
where we introduce, for simplicity, a dimensionless parameter $\epsilon_{24} \equiv v_{24}/\Lambda$.   Our notation is such that $M_E$ represents mass matrix for charged leptons, $M_D$ is the down-type quark mass matrix, and $M_U$ is the up-type quark mass matrix.  Moreover, these mass matrices are written in the $f^Cf$ basis, where $f$ stands for the appropriate SM charged fermions.

The ordering of the scales is such that $\Lambda > v_{24} \gg v_5$, where $v_5=246$\,GeV. The scale of $v_{24}$ is proportional to the masses of the proton mediating gauge bosons via \begin{align}
M_{X,Y}=\sqrt{25/8} g_\mathrm{GUT} v_{24},
\end{align}
where $g_\mathrm{GUT}$ is a gauge coupling constant of $SU(5)$ at the scale of unification. The $X$ and $Y$ gauge boson mass $M_{X,Y}$, on the other hand, can be identified with the scale of gauge coupling unification $M_\mathrm{GUT}$ and originates from kinetic term in the lagrangian
\begin{align}
&\mathcal{L}_K= \frac{1}{2} \left(D_\mu \Phi\right)^* \left(D^\mu \Phi\right),    
\end{align}
where $\Phi^i_j \equiv 24_H$ and $D_\mu \Phi^i_j= \partial_\mu \Phi^i_j+ ig_\mathrm{GUT} \left(A_\mu\right)^m_j \Phi^i_m -ig_\mathrm{GUT} \left(A_\mu\right)^i_n \Phi^n_j$.

Since we often spell out our results in the physical basis for the SM fermions, we specify, for definiteness, that the transition between the flavor basis and the mass eigenstate basis is implemented through the following set of transformations:
\begin{align}
&E^T_c M_E E=  M_E^\mathrm{diag}, \label{ME}\\ 
&D^T_c M_D D=  M_D^\mathrm{diag},  \label{MD}\\
&U^T_c M_U U=  M_U^\mathrm{diag}, \label{MU}\\
&N^T M_N N=  M_N^\mathrm{diag}. \label{MN}
\end{align}
Here $E_c$, $E$, $D_c$, $D$, $U_c$, $U$, and $N$ are {\it a priori} arbitrary $3 \times 3$ unitary matrices. $M_N \equiv M^T_N$ is a $3 \times 3$ mass matrix for neutrinos, where we assume neutrinos to be of Majorana nature. 

\subsection{Color-triplet couplings}

The couplings of the triplet $T(3,1,-1/3) \in 5_H$ to the SM fermions in the mass eigenstate basis, as given by lagrangian of Eq.\ \eqref{eq:lagrangian_24} and conventions presented in Eqs.\ \eqref{ME} though \eqref{MN}, are~\cite{Dorsner:2024seb}\\
$(i)\;u^T_{k,\alpha}C^{-1}e_\beta T^*_k:$
\begin{align}
    -\frac{1}{\sqrt{2}}\bigg\{U^T
\bigg[
Y_d- Y_1 \epsilon_{24} -\frac{3}{2} Y_2 \epsilon_{24}
\bigg] E
\bigg\}_{\alpha\beta},  \label{scalar-1}
\end{align}
\noindent 
$(ii)\;d^T_{k,\alpha}C^{-1}\nu_\beta T^*_k:$
\begin{align}
    \frac{1}{\sqrt{2}}\bigg\{D^T
\bigg[
Y_d- Y_1 \epsilon_{24} -\frac{3}{2} Y_2  \epsilon_{24}
\bigg] N
\bigg\}_{\alpha\beta},  \label{scalar-2}
\end{align}
\noindent
$(iii)\;\epsilon_{ijk} u^{C,T}_{i,\alpha}C^{-1}d^C_{j,\beta} T^*_k:$
\begin{align}
    \frac{1}{\sqrt{2}}\bigg\{U_c^\dagger
\bigg[Y_d - Y_1 \epsilon_{24} +Y_2  \epsilon_{24}
\bigg] D_c^*
\bigg\}_{\alpha\beta},  \label{scalar-3}
\end{align}
\noindent
$(iv)\;\epsilon_{ijk} u^T_{i,\alpha}C^{-1}d_{j,\beta} T_k:$
\begin{align}
    -\bigg\{U^T
\bigg[
2\left(Y_u+Y_u^T\right) -2\left(Y_3+Y_3^T\right)\epsilon_{24}
+\frac{1}{2} \left(Y_4+Y_4^T\right)\epsilon_{24}
\bigg] D
\bigg\}_{\alpha\beta},  \label{scalar-4}
\end{align}
\noindent
$(v)\;u^{C,T}_{k,\alpha}C^{-1}e^C_\beta T_k:$
\begin{align}
    \bigg\{U^\dagger_c
\bigg[
2\left(Y_u+Y_u^T\right) -2\left(Y_3+Y_3^T\right)\epsilon_{24}
+\left(3Y_4-2Y^T_4\right)\epsilon_{24}
\bigg] E_c^*
\bigg\}_{\alpha\beta}, \label{scalar-5}
\end{align}
where Eqs.\ \eqref{scalar-1}, \eqref{scalar-2}, and \eqref{scalar-3} originate from $SU(5)$ contractions between $10_F^\alpha$ and $\overline{5}_{F}^\beta$, whereas Eqs.\ \eqref{scalar-4} and \eqref{scalar-5} originate from contractions between $10_F^\alpha$ and $10_F^\beta$. Note that we use $T_i \equiv 5_H^i$ and $T^*_i \equiv 5^*_{Hi}$, where $i=1,2,3$.

The triplet $T$ couples simultaneously to the quark-lepton and quark-quark pairs at both the $d=4$ and $d=5$ levels. It is, nevertheless,  possible to suppress either quark-quark or quark-lepton couplings of the triplet and thus prevent tree-level two-body proton decay due to the triplet mediation through implementation of specific relations between $Y_d$, $Y_1$, $Y_2$, $Y_u$, $Y_3$, and $Y_4$~\cite{Dorsner:2024seb}. 

For example, the quark-quark pair interactions with the triplet $T$ can be completely suppressed with the following two conditions
\begin{align}
  Y_d-Y_1\epsilon_{24}+ Y_2\epsilon_{24}&=0,
\label{eq:qq_a_24}\\
 \left(Y_u+Y_u^T\right)-(Y_3+Y_3^T)  \epsilon_{24}+\frac{1}{4} \left(Y_4+Y_4^T\right)\epsilon_{24}   &=0.
\label{eq:qq_b_24}
\end{align}
The suppression of the quark-lepton pair interactions with the triplet, on the other hand, can be accomplished via
\begin{align}
Y_d-Y_1\epsilon_{24}+\frac{3}{2} Y_2\epsilon_{24}&=0,
\label{eq:ql_a_24}\\
\left(Y_u+Y_u^T\right)-(Y_3+Y_3^T)  \epsilon_{24}+ \left(\frac{3}{2}Y_4-Y_4^T\right)\epsilon_{24}   &=0
\label{eq:ql_b_24}.
\end{align}
Even though Eqs.\ \eqref{eq:qq_a_24} and \eqref{eq:qq_b_24} or Eqs.\ \eqref{eq:ql_a_24} and \eqref{eq:ql_b_24} impose certain constraints on the particular form of Yukawa coupling matrices, these constraints are not in conflict with viable generation of charged fermion masses. With this in mind, several additional observations are in order. 

Firstly, if one is to completely suppress either the quark-quark or quark-lepton couplings of the triplet $T \in 5_H$ in a phenomenologically viable manner, one needs all three contractions between $10_F^\alpha$ and $\overline{5}_{F}^\beta$ that are featured in the first line of Eq.\ \eqref{eq:lagrangian_24}. The reason for that is very simple. Namely, one needs to simultaneously suppress either the quark-lepton or quark-quark  interactions of the triplet while still generating experimentally observed masses of charged leptons and down-type quarks via $M_E$ and $M_D$ mass matrices, respectively. 

The up-type quark sector is much less demanding since a viable $M_U$ can be successfully generated with the first and/or second contribution in Eq.\ \eqref{eq:uncorrected_U_24}. Moreover, there is a fortuitous alignment in Eqs.\ \eqref{scalar-4} and~\eqref{scalar-5} between contributions proportional to $Y_u$ and $Y_3$. One can thus simultaneously suppress both quark-lepton and quark-quark couplings of the triplet, while maintaining viability of $M_U$ with the presence of only the first two contractions between $10_F^\alpha$ and $10_F^\beta$ that are featured in the second line of Eq.\ \eqref{eq:lagrangian_24}, if needed. Again, even if $Y_4$ is taken to be a null-matrix, one can set to zero the triplet interactions in Eqs.\ \eqref{scalar-4} and \eqref{scalar-5} and still be able to produce viable mass matrix for the up-type quarks.   

The level of required cancellation between different Yukawa couplings, in order to suppress either the quark-lepton or quark-quark interactions of the triplet, can be expressed in terms of the triplet mass. Namely, if we denote the triplet mass with $m_T$, the level of cancellation between corresponding matrix elements needs to be at the level of $m_T/(10^{12}\,\mathrm{GeV})$~\cite{Dorsner:2012uz,Dorsner:2024jiy}. Moreover, this cancellation needs to take place at the scale relevant for the proton decay experiments since the relations presented in Eqs.\ \eqref{eq:qq_a_24} and \eqref{eq:qq_b_24} or Eqs.\ \eqref{eq:ql_a_24} and \eqref{eq:ql_b_24} are certainly not invariant under the renormalization group equation running. Also, these relations are not result of some particular symmetry that could provide deeper understanding of their origin.

Finally, what we are advocating is a potential suppression of the triplet couplings in an arbitrary flavor basis as the unitary transformations $E_c$, $E$, $D_c$, $D$, $U_c$, $U$, and $N$ need not be specified at all. This simply means that there are infinitely many ways to implement the suppression of interest. Also, one can add $d>5$ terms to Eq.\ \eqref{eq:lagrangian_24} to introduce even more parameter freedom to the problem, if needed, and/or resort to a use of unitary transformations to aid with suppression of the proton decay inducing interactions. Be that as it may, our discussion demonstrates that it is entirely possible to bypass the experimental source of the doublet-triplet splitting problem. Simply put, our approach provides a light triplet that can still couple to the SM fermions as long as one introduces higher-dimensional $SU(5)$ contractions. What one minimally needs to accomplish the suppression are three distinct $SU(5)$ contractions between $10_F^\alpha$ and $\overline{5}_{F}^\beta$ and two contractions between $10_F^\alpha$ and $10_F^\beta$.

\begin{figure}[th!]
\centering
\includegraphics[width=0.6\linewidth]{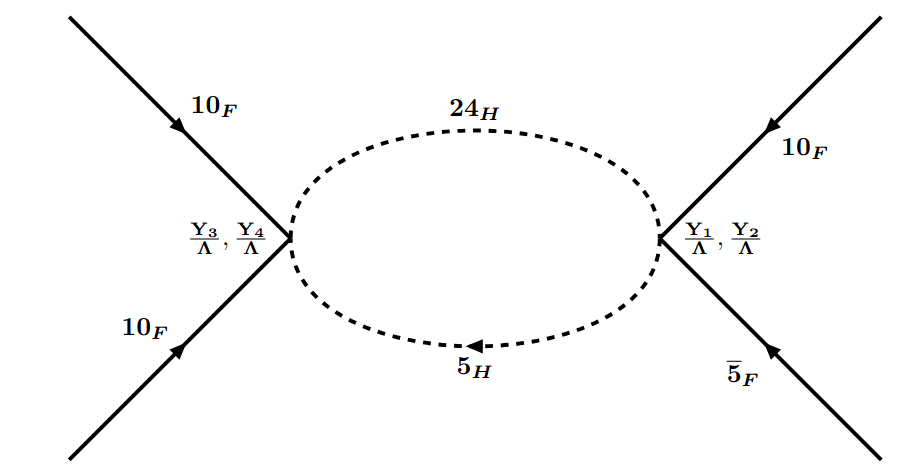}
\caption{A one-loop level proton decay inducing diagram that utilizes $d=5$ operators at each vertex, as indicated.}
\label{fig:loopPD}
\end{figure}

One can ask whether a complete suppression of the tree-level proton decay signatures induced by the triplet exchange is potentially violated at the loop level. What we have in mind is a type of process that is shown in Fig.\ \ref{fig:loopPD}, where the scalars in the loop reside in $24_H$ and $5_H$. 

To answer this question we observe that one vertex of the proton decay inducing diagram of Fig.\ \ref{fig:loopPD} must originate, due to group theoretical reasons, from the contraction of $10_F^\alpha$ with $10_F^\beta$, whereas the other vertex corresponds to a contraction of $10_F^\alpha$ with $\overline{5}_{F}^\beta$. The maximal value of the Yukawa coupling(s) at the $10_F^\alpha$-$\overline{5}_{F}^\beta$ vertex should always be suppressed with respect to the corresponding maximal Yukawa coupling value at the $10_F^\alpha$-$10_F^\beta$ vertex to reflect observed mass hierarchy as there is only one electroweak VEV present. Also, both of these vertices are of at least the $d=5$ origin, as indicated in Fig.\ \ref{fig:loopPD}, and are thus inversely proportional to the cutoff scale $\Lambda$. It is then the largeness of $\Lambda$ and the usual loop suppression factor that make this contribution towards proton decay negligible even if one assumes order one Yukawa coupling entries in $Y_3$ and $Y_4$. In fact, the relevant Yukawa couplings are actually rather small as they are related to the SM fermion masses through Eqs.\ \eqref{eq:qq_a_24} and \eqref{eq:qq_b_24} or Eqs.\ \eqref{eq:ql_a_24} and \eqref{eq:ql_b_24}, depending on the suppression scenario at play.     

We have, so far, explicitly assumed that the state $\Phi_2(1,3,0) \in 24_H$ does not get a VEV. If that is not the case, the inclusion of its VEV in Eq.\ \eqref{eq:24vev} in the form of
\begin{align}
\langle 24_H \rangle & = \mathrm{diag}\left( -v_{24}, -v_{24}, -v_{24}, 3/2v_{24}+v_{3}, 3/2v_{24}-v_{3} \right) 
\end{align}
yields the following additional interaction terms between $T_i$ and the SM fermions:
\begin{align}
(i)\;u^T_{k,\alpha}C^{-1}e_\beta T^*_k:&\,\,\,
\bigg\{U^T
\bigg[
\frac{1}{\sqrt{2}} Y_2 \epsilon_{3}
\bigg] E
\bigg\}_{\alpha\beta},  \label{scalar-1New}\\
(ii)\;d^T_{k,\alpha}C^{-1}\nu_\beta T^*_k:&\,\,\,
\bigg\{D^T
\bigg[
\frac{1}{\sqrt{2}} Y_2  \epsilon_{3}
\bigg] N
\bigg\}_{\alpha\beta},  \label{scalar-2New}\\
(iii)\;\epsilon_{ijk} u^{C,T}_{i,\alpha}C^{-1}d^C_{j,\beta} T^*_k:&\,\,\,
0,  \label{scalar-3New}\\
(iv)\;\epsilon_{ijk} u^T_{i,\alpha}C^{-1}d_{j,\beta} T_k:&\,\,\,
-\bigg\{U^T
\bigg[
\left(Y_4-Y_4^T\right)\epsilon_{3}
\bigg] D
\bigg\}_{\alpha\beta},  \label{scalar-4New}\\
(v)\;u^{C,T}_{k,\alpha}C^{-1}e^C_\beta T_k:&\,\,\,
0. \label{scalar-5New}
\end{align}
Here, we introduce another dimensionless parameter $\epsilon_{3}\equiv v_{3}/\Lambda$. Clearly, the $SU(5)$ symmetry and a particular direction of the VEV that is proportional to $v_3$ dictates an absence of several interaction terms between the SM fermions and the triplet that would otherwise be allowed by the SM gauge group symmetry. But, even if both quark-quark and quark-lepton interactions are simultaneously present, the electroweak precision measurements place a stringent upper limit on the value of $v_3(<v_5)$. This, on the other hand, stipulates that terms proportional to $\epsilon_{3}$ can be safely neglected for all practical purposes when considering impact on proton stability. 

\section{75-Higgs}
\label{sec:75}

The choice of the scalar representation that breaks $SU(5)$ gauge group is not unique even if the phenomenologically viable symmetry breaking chain $SU(5) \rightarrow SU(3) \times SU(2) \times U(1) \rightarrow SU(3) \times U(1)_\mathrm{em}$ is. Namely, one can accomplish aforementioned breaking of $SU(5)$ by using a $75$-dimensional representation instead of a $24$-dimensional one~\cite{Hubsch:1984pg}, where the decomposition of $75_H$ under the SM gauge group is
\begin{align}75_H &= \Phi_1 (1,1,0) +
\Phi_2 (8,1,0) +
\Phi_3 (8,3,0) +
\Phi_4 (3,1,5/3) 
+
\Phi_4^* (\overline{3},1,-5/3)
\nonumber\\& +
\Phi_5 (3,2,-5/6) 
+
\Phi_5^* (\overline{3},2,5/6)
+
\Phi_6 (\overline{6},2,-5/6) 
+
\Phi_{6}^* (6,2,5/6).
\end{align} We accordingly investigate if it is possible to implement suppression of the triplet interactions with the SM fermions within the $75$-dimensional scenario and if it defers from the $24$-dimensional scenario in that regard. Of course, an obvious difference is that $75$-dimensional representation has only one state that can get phenomenologically viable VEV, whereas $24$-dimensional representation has two such states. In other words, the proton decay inducing couplings of the sort presented in Eqs.\ \eqref{scalar-1New} through \eqref{scalar-5New} simply do not exist within the $75$-dimensional scenario.

The symmetry properties of $75_H\equiv\Phi^{ij}_{kl}$ are $\Phi^{ij}_{kl}=-\Phi^{ji}_{kl}=-\Phi^{ij}_{lk}=+\Phi^{ji}_{lk}$ and $\sum_{i=1}^5 \Phi^{ij}_{il}=0$, where $i,j,k,l=1,\ldots,5$ are, once again, $SU(5)$ indices. The VEV structure of $75_H$ that breaks $SU(5)$ down to $SU(3) \times SU(2) \times U(1)$ can be summarised as follows
\begin{align}
\nonumber
\langle 75_H \rangle &=(\Phi^{12}_{12},\Phi^{13}_{13},\Phi^{23}_{23},\Phi^{14}_{14},\Phi^{15}_{15},\Phi^{24}_{24},\Phi^{25}_{25},\Phi^{34}_{34},\Phi^{35}_{35},\Phi^{45}_{45})
\\
&=\frac{v_{75}}{3\sqrt{2}} (1,1,1,-1,-1,-1,-1,-1,-1,3).     
\end{align}
We furthermore assume that $5_H$ breaks $SU(3) \times SU(2) \times U(1)$ down to $SU(3) \times U(1)_\mathrm{em}$ with the VEV of Eq.\ \eqref{eq:5vev}.

The masses of proton decay mediating gauge bosons $X$ and $Y$, in this scenario, are given by 
\begin{align}
M_{X,Y}= \sqrt{8/3}g_\mathrm{GUT} v_{75},
\end{align}
where $g_\mathrm{GUT}$, once again, is a gauge coupling constant of $SU(5)$ at the scale of unification. The kinetic term in the Lagrangian that yields $M_{X,Y}^2$ is 
\begin{align}
&\mathcal{L}_K= \frac{1}{2} \left(D_\mu \Phi\right)^* \left(D^\mu \Phi\right),    
\end{align}
where $D_\mu \Phi^{ij}_{kl}= \partial_\mu \Phi^{ij}_{kl}+ ig_\mathrm{GUT} \left( \left(A_\mu\right)^i_m \Phi^{mj}_{kl} + \left(A_\mu\right)^j_n \Phi^{in}_{kl} -  \left(A_\mu\right)^p_k \Phi^{ij}_{pl} -  \left(A_\mu\right)^q_l \Phi^{ij}_{kq}\right)$.

\subsection{Yukawa couplings}

The Yukawa interactions responsible for generating the charged fermion masses are 
\begin{align}
&\mathcal{L}_Y= 
10_F^{\alpha ij} \overline{5}_{Fk}^\beta 5^*_{Hl} \bigg\{
Y_{a\alpha\beta} \delta_{i}^k\delta_{j}^l   +
\frac{1}{\Lambda} Y_{b\alpha\beta}   75_{Hij}^{kl}
+
\frac{1}{\Lambda^2} Y_{c\alpha\beta}     75^{km}_{Hin}75^{ln}_{Hjm}
+
\frac{1}{\Lambda^2} Y_{d\alpha\beta}     75^{mn}_{Hij}75^{kl}_{Hmn}
\bigg\}
\nonumber\\&+
10_F^{\alpha ij}10_F^{\beta kl} 5_H^m
\bigg\{
Y_{A\alpha\beta}   \epsilon_{ijklm}
+ 
\frac{1}{\Lambda}Y_{B\alpha\beta}   \epsilon_{ijnom} 75^{no}_{Hkl}
+
\frac{1}{\Lambda}Y_{C\alpha\beta}   \epsilon_{jklno} 75^{no}_{Him}+
\frac{1}{\Lambda}Y_{D\alpha\beta}   \epsilon_{jlmno} 75^{no}_{Hik}\bigg\}
\nonumber\\&+\mathrm{h.c.},
\label{eq:lagrangian_75}
\end{align}
where we include all possible $d=4$, $d=5$, and $d=6$ contractions between $10_F^\alpha$ and $\overline{5}_{F}^\beta$. The reason behind inclusion of all these contractions will be discussed in detail later on. 

\subsection{Mass matrices}
The mass matrices of the SM charged fermions, as given by Eq.\ \eqref{eq:lagrangian_75}, are
\begin{align}
M_E &=v_5 \bigg\{\frac{1}{2}Y_a +\frac{1}{\sqrt{2}} Y_b \epsilon_{75}  - \frac{1}{6} Y_c \epsilon^2_{75}  + Y_d \epsilon^2_{75} \bigg\},\label{eq:uncorrected_E_75}\\
M_D &=v_5 \bigg\{\frac{1}{2}Y_a^T -\frac{1}{3\sqrt{2}} Y_b^T \epsilon_{75} -\frac{1}{6} Y_c^T \epsilon^2_{75}  + \frac{1}{9} Y_d^T \epsilon^2_{75}  \bigg\}, \label{eq:uncorrected_D_75}\\
M_U &=v_5 \bigg\{  \sqrt{2}\left(Y_A+Y_A^T\right) -\frac{2}{3}\left(Y_B-Y_B^T\right)\epsilon_{75}+\frac{2}{3}\left(Y_C-Y_C^T\right)\epsilon_{75} \bigg\},\label{eq:uncorrected_U_75}
\end{align}
where we introduce a dimensionless parameter $\epsilon_{75} =v_{75}/\Lambda$. The ordering of relevant scales is such that $\Lambda > v_{75} \gg v_5$. To go to the mass eigenstate basis for the SM charged fermions, i.e., to go from $M_{E,D,U}$ to $M^\mathrm{diag}_{E,D,U}$, one would need to perform unitary transformations introduced in Eqs.\ \eqref{ME}, \eqref{MD}, and \eqref{MU}. 

\subsection{Color-triplet couplings}

The triplet $T \in 5_H$ interactions with the SM fermions, as derived from Eq.\ \eqref{eq:lagrangian_75}, are\\
$(i)\;u^T_{k,\alpha}C^{-1}e_\beta T^*_k:$
\begin{align}
-\bigg\{U^T
\bigg[
\frac{Y_a}{\sqrt{2}}- \frac{1}{3} Y_b \epsilon_{75} -\frac{1}{3\sqrt{2}} Y_c \epsilon^2_{75}
 + \frac{\sqrt{2}}{9} Y_d \epsilon^2_{75}  
\bigg] E
\bigg\}_{\alpha\beta},  \label{scalar-1-75}
\end{align}
\noindent $(ii)\;d^T_{k,\alpha}C^{-1}\nu_\beta T^*_k:$
\begin{align}
\bigg\{D^T
\bigg[
\frac{Y_a}{\sqrt{2}}-\frac{1}{3} Y_b \epsilon_{75} -\frac{1}{3\sqrt{2}} Y_c  \epsilon^2_{75}
 + \frac{\sqrt{2}}{9} Y_d \epsilon^2_{75}  
\bigg] N
\bigg\}_{\alpha\beta},  \label{scalar-3-75}
\end{align}
\noindent $(iii)\;\epsilon_{ijk} u^{C,T}_{i,\alpha}C^{-1}d^C_{j,\beta} T^*_k:$
\begin{align}
\bigg\{U_c^\dagger
\bigg[\frac{Y_a}{\sqrt{2}}+\frac{1}{3} Y_b \epsilon_{75} +\frac{1}{9\sqrt{2}} Y_c  \epsilon^2_{75}
 + \frac{\sqrt{2}}{9} Y_d \epsilon^2_{75} 
\bigg] D_c^*
\bigg\}_{\alpha\beta},  \label{scalar-2-75}
\end{align}
\noindent $(iv)\;\epsilon_{ijk} u^T_{i,\alpha}C^{-1}d_{j,\beta} T_k:$
\begin{align}
\bigg\{U^T
\bigg[
-2\left(Y_A+Y_A^T\right) +\frac{2\sqrt{2}}{3}\left(Y_B+Y_B^T\right)\epsilon_{75}+\frac{\sqrt{2}}{3}\left(Y_D+Y_D^T\right)\epsilon_{75}
\bigg] D
\bigg\}_{\alpha\beta},  \label{scalar-4-75}
\end{align}
\noindent $(v)\;u^{C,T}_{k,\alpha}C^{-1}e^C_\beta T_k:$
\begin{align} 
\bigg\{U^\dagger_c
\bigg[
2\left(Y_A+Y_A^T\right) +\frac{\sqrt{8}}{3}\left(3 Y_B+Y_B^T\right)\epsilon_{75}-\frac{ \sqrt{8}}{3}\left(Y_C-Y_C^T\right)\epsilon_{75}+\frac{ \sqrt{8}}{3}\left(Y_D+Y_D^T\right)\epsilon_{75}
\bigg] E_c^*
\bigg\}_{\alpha\beta}. \label{scalar-5-75}
\end{align}

It is clear that the triplet interactions with the quark-quark pairs can be completely suppressed with the following two conditions
\begin{align}
  \frac{Y_a}{\sqrt{2}}+\frac{1}{3} Y_b \epsilon_{75} +\frac{1}{9\sqrt{2}} Y_c  \epsilon^2_{75}
+ \frac{\sqrt{2}}{9} Y_d \epsilon^2_{75}&=0,
\label{eq:qq_a_75}\\
 -\left(Y_A+Y_A^T\right) +\frac{\sqrt{2}}{3}\left(Y_B+Y_B^T\right)\epsilon_{75}+\frac{\sqrt{2}}{6}\left(Y_D+Y_D^T\right)\epsilon_{75}&=0,
\label{eq:qq_b_75}
\end{align}
whereas the quark-lepton-leptoquark interactions can be set to zero via
\begin{align}
\frac{Y_a}{\sqrt{2}}- \frac{1}{3} Y_b \epsilon_{75} -\frac{1}{3\sqrt{2}} Y_c \epsilon^2_{75} + \frac{\sqrt{2}}{9} Y_d \epsilon^2_{75}&=0,
\label{eq:ql_a_75}\\
\left(Y_A+Y_A^T\right) +\frac{\sqrt{2}}{3}\left(3 Y_B+Y_B^T\right)\epsilon_{75}-\frac{ \sqrt{2}}{3}\left(Y_C-Y_C^T\right)\epsilon_{75}+\frac{ \sqrt{2}}{3}\left(Y_D+Y_D^T\right)\epsilon_{75}&=0\label{eq:ql_b_75}.
\end{align}

There are several crucial differences between the $75$-dimensional and $24$-dimensional symmetry breaking scenarios when it comes to the generation of the SM charged fermion masses and the associated interactions with the color triplet as we discuss next. 

First, there is only one $d=5$ contraction that couples $10_F^\alpha$ to $\overline{5}_F^\beta$ in the $75$-dimensional scenario. One accordingly needs to introduce $d=6$ contraction(s) in Eq.\ \eqref{eq:lagrangian_75} to be able to simultaneously introduce viable down-type quark and charged lepton mass matrices and still be able to forbid the triplet couplings of either quark-quark or quark-lepton nature. Simply put, the $75$-dimensional scenario requires at least one $d=6$ contraction between $10_F^\alpha$ and $\overline{5}_F^\beta$ if one is to suppress proton decay inducing triplet couplings.  
Second, there are three possible $d=5$ terms that couple $10_F^\alpha$ and $10_F^\beta$, as can be seen from Eq.\ \eqref{eq:lagrangian_75}. This means that it is trivial to simultaneously address viable generation of the up-type quark masses and suppress proton decay inducing interactions of the triplet with the SM fermions that are associated with the $10_F^\alpha$-$10_F^\beta$ contractions. Moreover, the $SU(5)$ contraction featuring $Y_D$ in Eq.\ \eqref{eq:lagrangian_75} generates interactions between the SM fermions and the triplet but does not generate any contribution towards the up-type quark masses. This means that it is even possible to suppress interactions between the triplet and the SM fermions without imposing any conditions on the flavor structure of the up-type quark mass matrix.  

We can conclude that the $75$-dimensional scenario also allows for a light color triplet scalar as long as one includes higher-dimensional operators in the Yukawa sector of the theory. One prominent feature to remember is that the $75$-dimensional scenario requires at least one $d=6$ contraction between $10_F^\alpha$ and $\overline{5}_F^\beta$ to be present if one is to have a light triplet. 

\section{Leptoquark-Induced Neutrino Masses}
\label{sec:neutrinos}

One can ask what new model building avenues can be accessible in view of the fact that the  proton decay inducing interactions of the color triplet might be suppressed if one allows introduction of higher-dimensional operators into the theory. 

To answer that question we first investigate viability of two simple extensions of the Georgi-Glashow model, in the light triplet regime, that can generate phenomenologically viable masses of all SM fermions. One extension requires a presence of a single $10$-dimensional scalar representation, whereas the other one relies on an addition of a single $15$-dimensional scalar representation, where, in both instances, neutrino masses are taken to be of the one-loop~\cite{Zee:1980ai} level origin. 

We subsequently replace a $24$-dimensional representation with a $75$-dimensional representation and proceed to investigate viability of the one-loop level neutrino mass generation within a $10$-dimensional and a $15$-dimensional scalar representation extensions of aforementioned symmetry breaking scenarios.  Again, we are solely interested in a regime when the triplet $T \in 5_H$ is light since that particular limit has not been discussed in the literature. (For neutrino mass generation via loops within the $SU(5)$ framework, see, for example, Refs.~\cite{Wolfenstein:1980sf,Barbieri:1981yw,Perez:2016qbo,Dorsner:2017wwn,Kumericki:2017sfc,Saad:2019vjo,Dorsner:2019vgf,Dorsner:2021qwg,Antusch:2023jok,Dorsner:2024jiy,Klein:2019jgb,Hinze:2024vrl,Babu:2024jdw,Dogan:2025stk}. For other related works, see also Refs.~\cite{Chang:1980ey,Babu:2001ex,Babu:2010vp,deGouvea:2014lva,Boucenna:2014dia,Hagedorn:2016dze,Cai:2017jrq}.)

\subsection{The $24_H$ scenario case studies}

\subsubsection{Extension with a 10-dimensional scalar representation}

\begin{figure}[th!]
\centering
\includegraphics[width=0.7\linewidth]{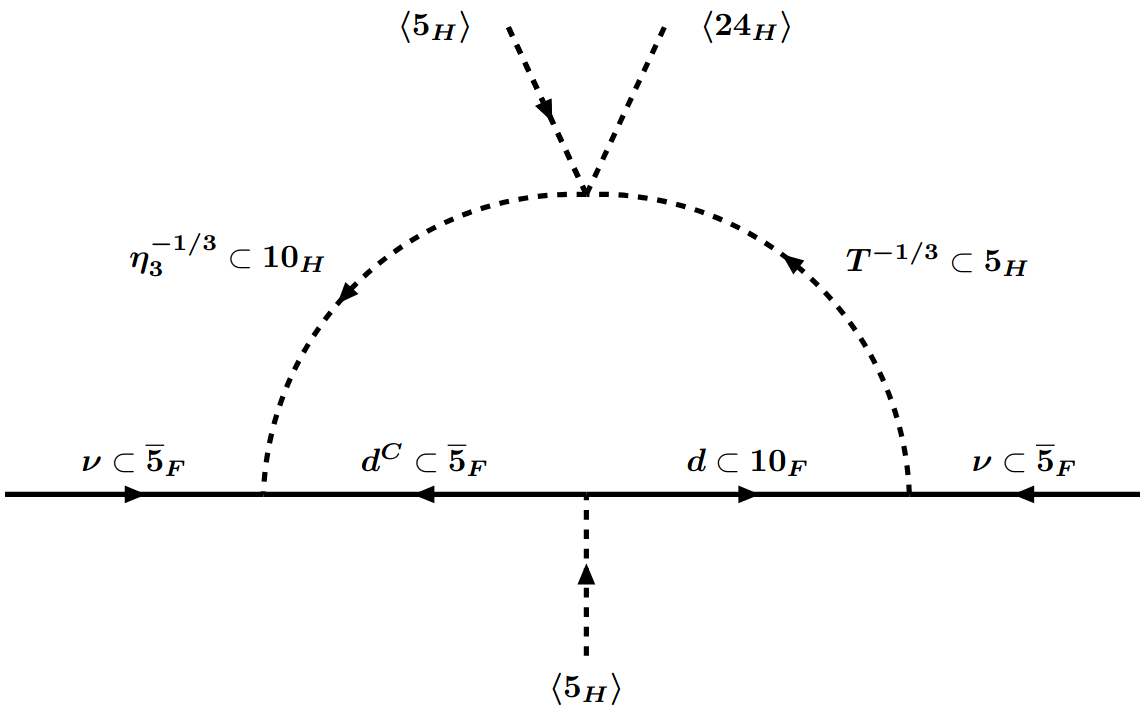}
\caption{One-loop neutrino mass generating diagram within the $10_H$ extension, when the $SU(5)$ symmetry breaking is accomplished with representations  $24_H$ and $5_H$.}
\label{fig:Nu10}
\end{figure}

If the Georgi-Glashow model is extended with a single $10$-dimensional scalar representation $10_H$, one can generate neutrino masses at the one-loop level through a diagram that is shown in Fig.\ \ref{fig:Nu10}. 

To complete the loop of Fig.\ \ref{fig:Nu10} one needs an interaction term between $10_H$ and the SM fermions as well as the mixing term between relevant leptoquarks in $10_H$ and $5_H$. Recall, the decomposition of $10_H$, under the SM gauge group $SU(3)\times SU(2)\times U(1)$, is
\begin{equation}
10_H= \eta_1 (1,1,1)+\eta_2 (\overline{3},1,-2/3)+ \eta_3 (3,2,1/6),
\end{equation}
where $\eta_3^{-1/3} \in \eta_3 (3,2,1/6)$ is one of the leptoquarks in question. The other leptoquark is, of course, the color triplet $T^{-1/3} \in 5_H$. Note that we use superscripts to explicitly denote electric charges of leptoquarks in units of the positron charge.  

The Yukawa interactions of interest originate from  
\begin{align}
   -\mathcal{L}_Y \supset Y_{Y\alpha \beta} \overline 5_{F i}^\alpha \overline 5_{F j}^\beta 10_H^{ij}  +\frac{1}{\Lambda}  Y_{Z\alpha \beta} \overline 5_{F i}^\alpha \overline 5_{F j}^\beta 10_H^{ik} 24^{j}_{H k}
    \supset   \nu^T_\alpha C^{-1} Y_{X\alpha \beta}  d^C_\beta \eta_3^{-1/3},
\label{eq:numass10H}
\end{align}
where we also include one specific $d=5$ contraction. We will show later on that the inclusion of the $d=5$ term is essential for viable generation of neutrino masses and mixing parameters.   Note that $Y_X$ is defined in the flavor basis of the SM fermions and it reads  
\begin{align}
 Y_X=  \sqrt{2}Y_{Y} -   \frac{5}{2\sqrt{2}} 
Y_{Z}\epsilon_{24},
\label{eq:Y_X}
\end{align}
where $Y_Y$ is a skew-symmetric matrix in the flavor space, whereas $Y_Z$ is an arbitrary matrix. 

The relevant mixing between the scalar leptoquarks $T^{-1/3} \in 5_H$ and $\eta_3^{-1/3} \in 10_H$, necessary for generating neutrino masses, arises from the following term  
\begin{align}
    V \supset  \lambda \; 5^*_{H i} 5^*_{H j} 10_H^{ik} 24^j_{H k} \supset \frac{5}{4}\lambda v_5 v_{24} T^{1/3} \eta^{-1/3}_3,
    \label{eq:mixing10}
\end{align}
where we use $(T^{-1/3})^*=T^{1/3}$ for convenience.
Note that the cubic term $5^*_{H i} 5^*_{H j} 10_H^{ij} $ vanishes due to the skew-symmetric property of $10_H$ in the $SU(5)$ space.  

The mass-squared matrix for scalar leptoquarks reads
\begin{align}
\label{eq:msquared10}
M^2_S=    \begin{pmatrix}
m^2_T&\frac{5}{4}\lambda v_5 v_{24}
\\
\frac{5}{4}\lambda v_5 v_{24}&m^2_{\eta_3}
\end{pmatrix},
\end{align}
where $m_T$ and $m_{\eta_3}$ would be masses of $T^{-1/3}$ and $\eta_3^{-1/3}$, respectively, in the absence of the mixing term given in Eq.\ \eqref{eq:mixing10}.
If we introduce the mass eigenstates $S_1$ an $S_2$ for two scalar leptoquarks $T^{-1/3}$ and $\eta_3^{-1/3}$ via
\begin{align}
\label{eq:linear_comb}
\begin{pmatrix}
T^{-1/3}\\\eta_3^{-1/3}
\end{pmatrix}
=
\begin{pmatrix}
c_\theta&-s_\theta
\\
s_\theta&c_\theta
\end{pmatrix}
\begin{pmatrix}
S_1\\S_2
\end{pmatrix},
\end{align}
where $\theta$ takes $M^2_S$ of Eq.\ \eqref{eq:msquared10} into a diagonal form via
\begin{align}
    \tan 2\theta=\frac{ 5\lambda v_5 v_{24}/2 }{m^2_T-m^2_{\eta_3}},
\end{align}
the neutrino mass matrix of Fig.~\ref{fig:Nu10} reads~\cite{Perez:2016qbo}
\begin{align}
\label{eq:numass}
M_N \equiv M_N^T \approx \frac{3\sin2\theta}{32\pi^2} \ln\left( \frac{m_{S_1}^2}{m_{S_2}^2} \right) \bigg\{
Y_X D_c M_D^\mathrm{diag} D^T Y_T + Y_T^T D M_D^\mathrm{diag} D_c^T Y_X^T
\bigg\}.
\end{align}
Here, $Y_T$ is the Yukawa coupling matrix of $T \in 5_H$ with the $d$-$\nu$ pairs in the flavor basis that can be taken directly from Eq.\ \eqref{scalar-2}, where one should omit unitary transformations of the SM fermions while $m_{S_1}$ and $m_{S_2}$ are masses of scalars $S_1$ and $S_2$, respectively. Note that the Pontecorvo-Maki-Nakagawa-Sakata (PMNS) mixing matrix is defined to be $U_\mathrm{PMNS}= E^\dagger N$, where $M_N= N^*   M^\mathrm{diag}_N    N^\dagger$, in agreement with Eq.\ \eqref{MN}. 

The neutrino masses in Eq.\ \eqref{eq:numass} vanish for exact mass degeneracy between $S_1$ and $S_2$, i.e., when $m_{S_1}=m_{S_2}$. However, phenomenologically viable neutrino masses can be obtained even when $m_{S_1}\approx m_{S_2}$ for $\mathcal{O}(1)$  Yukawa couplings. 

To proceed, we need to address the question of gauge coupling unification within the model comprising $24_H$, $10_H$, $5_H$, $10_F^\alpha$, and $\overline{5}_{F}^\alpha$, where $\alpha=1,2,3$. To that end we implement one-loop level gauge coupling unification analysis in order to find the largest possible value of unification scale $M_\mathrm{GUT}$ and associated value of $g^2_\mathrm{GUT}=4 \pi \alpha_\mathrm{GUT}$ for the fixed values of $m_{S_1}$ and $m_{S_2}$, where we take $S_1$ and $S_2$ to be mass degenerate for simplicity. The relevant central values of the SM input parameters that we use for unification study are  $M_Z=91.1876$\,GeV, $\alpha_S(M_Z)=0.1193$, $\alpha^{-1}(M_Z)=127.906$, and
$\sin^2 \theta_W(M_Z)=0.23126$~\cite{Agashe:2014kda}.

It turns out that the unification does not take place within the $10_H$ extension of the Georgi-Glashow model unless one also takes into account higher-dimensional contributions towards kinetic terms for the gauge fields~\cite{Hill:1983xh,Shafi:1983gz,Chakrabortty:2008zk}  
\begin{align}
\mathcal{L}_5 \supset -\frac{c_5}{\Lambda} \bigg\{  \frac{1}{2} Tr\left( F_{\mu\nu}  24_H F^{\mu\nu}  \right) \bigg\}, 
\end{align}
where $c_5$ is a dimensionless parameter.
If we introduce another dimensionless parameter $\epsilon_5$ via
\begin{align}
& \epsilon_5=\frac{c_5 v_{24}}{2\Lambda},
\label{eq:epsilon_5}
\end{align}
the modified gauge coupling unification conditions, at $M_\mathrm{GUT}$ scale, become~\cite{Hill:1983xh,Shafi:1983gz,Chakrabortty:2008zk}
\begin{align}
g^2_1(M_\mathrm{GUT})(1+ \epsilon_5) 
=
g^2_2(M_\mathrm{GUT})(1+ 3\epsilon_5) 
=
g^2_3(M_\mathrm{GUT})(1-2 \epsilon_5). 
\end{align} 
This, then, allows for gauge coupling unification for judiciously chosen values of $\epsilon_5$.

We present the results of our gauge coupling unification analysis in Table~\ref{tab:unification10H}, where we provide the highest possible value of $M_\mathrm{GUT}$ as a function of $m_{S_1}=m_{S_2}$ as well as the associated values of $\epsilon_5$ and $\alpha_\mathrm{GUT}^{-1}$. The automated unification procedure looks for the highest possible unification scale $M^\mathrm{max}_\mathrm{GUT}$ by treating the masses of all other scalars in $24_H$ and $10_H$ to be free parameters that can take any value between $1$\,TeV and $M_\mathrm{GUT}$.  
\begin{table} 
\centering
\begin{tabular}{|c|c|c|c|}
  \hline
$\epsilon_5$ & $m_{S_1}=m_{S_2}$ (TeV) & $M^\mathrm{max}_\mathrm{GUT}$  (10$^{14}$\,GeV) & $\alpha_\mathrm{GUT}^{-1}$ \\   
\hline \hline 
$0.020$ & $10^0$ & $5.933$ & $38.2$ \\ 
$0.021$  & $10^1$ & $5.229$ & $38.3$ \\
$0.021$  & $10^2$ & $4.759$ & $38.4$ \\ 
$0.021$ & $10^3$ & $4.236$ & $38.5$ \\ 
$0.021$  & $10^4$ & $3.697$ & $38.6$ \\  
$0.022$  & $10^5$ & $3.338$ & $38.6$ \\  
  \hline
\end{tabular}
\caption{The highest possible unification scale $M^\mathrm{max}_\mathrm{GUT}$ as a function of degenerate masses of linear combinations of scalars $T^{-1/3}$ and $\eta_3^{-1/3}$ within the $10_H$ extension of the $24_H$ scenario.}
\label{tab:unification10H}
\end{table}

It is clear that the values for $M^\mathrm{max}_\mathrm{GUT}$ that are given in Table \ref{tab:unification10H} also require one to substantially suppress gauge mediated proton decay~\cite{Dorsner:2004xa}. This suppression places a set of constraints on potentially viable form of unitary matrices that are introduced in Eqs.\ \eqref{ME} through \eqref{MN}. The natural question then is whether one can simultaneously impose restrictions on the Yukawa coupling matrices in order to have a light triplet and restrict parameter space of unitary matrices in order to suppress gauge mediated proton decay and still be able to generate viable fermion masses. We address this question in detail in what follows.  

Firstly, the relevant interactions of the triplet with the SM fermions that enter $M_N$ of Eq.\ \eqref{eq:numass} are
\begin{align}
& Y_T= \frac{1}{\sqrt{2}}
Y_d- \frac{\epsilon_{24}}{\sqrt{2}} Y_1  -\frac{3\epsilon_{24}}{2\sqrt{2}} Y_2  .
\end{align}
Suppression of the triplet interactions with the quark-quark pairs and, consequentially, its proton decay signatures leads to
\begin{align}
\label{eq:spdsuppression}
&Y_d=\frac{4}{5v_5}M_E, \quad Y_1= \frac{4}{5v_5\epsilon_{24}}M_D^T, \quad Y_2= \frac{4}{5v_5\epsilon_{24}} \left(M_D^T-M_E \right).
\end{align}
This, in turn, yields
\begin{align}
Y_T=\frac{\sqrt{2}}{v_5}\left( M_E-M_D^T \right),
\label{eq:Y_T10_H}
\end{align}
and, consequentially, leads to
\begin{align}
M_N&=  a_0
\bigg\{
Y_X  D_c M_D^\mathrm{diag} D^T  \left(
E^*_c   M_E^\mathrm{diag} E^\dagger-D^*   M_D^\mathrm{diag} D_c^\dagger \right)
\nonumber\\&
+ \left( E^*   M_E^\mathrm{diag} E_c^\dagger - D^*_c   M_D^\mathrm{diag} D^\dagger  \right)  D M_D^\mathrm{diag} D^T_c  Y_X^T
\bigg\},
\end{align}
where we conveniently define
\begin{align}
    a_0=\frac{3\sqrt{2}\sin2\theta}{16\pi^2 v_5} \ln\left( \frac{m_{S_1}}{m_{S_2}} \right).
\end{align}

Secondly, the gauge mediated proton decay suppression~\cite{Dorsner:2004xa} is efficiently achieved if 
\begin{align}
&\left(  U^\dagger_c D  \right)_{1\alpha}=0, \quad \left(  E^\dagger_c D  \right)_{1\alpha}=\left(  E^\dagger_c D  \right)_{\alpha1}=0, \quad \left(  D^\dagger_c E  \right)_{1\alpha}=\left(  D^\dagger_c E  \right)_{\alpha1}=0,
\label{eq:pdsuppression}
\end{align}
where $\alpha=1,2$. (For the exact pattern of the two-body  proton decay signatures associated with the ansatz of Eq.\ \eqref{eq:pdsuppression} see Ref.\ \cite{Dorsner:2004xa}.) The most recent analysis~\cite{Senjanovic:2024uzn} of the impact of conditions in Eq.\ \eqref{eq:pdsuppression} on the lower bound on $M_\mathrm{GUT}$, in view of the current experimental limits on the partial proton decay lifetimes, quotes the following result
\begin{align}
 M_\mathrm{GUT} & \ge \sqrt{\alpha_\mathrm{GUT}/(40)^{-1}} 1.3 \times 10^{14}\,\mathrm{GeV}.  
\end{align} 
It is this limit that should be contrasted with the unification analysis results of Table \ref{tab:unification10H}.

The second and third condition of Eq.\ \eqref{eq:pdsuppression} translate to
\begin{align}
&E_c = D \begin{pmatrix}
0 &0&e^{i\xi_1}\\
0&e^{i\xi_2}&0\\
e^{i\xi_3}&0&0
\end{pmatrix}\equiv D P, 
\;\;\;
D_c = E \begin{pmatrix}
0 &0&e^{i\zeta_1}\\
0&e^{i\zeta_2}&0\\
e^{i\zeta_3}&0&0
\end{pmatrix}\equiv E Q
\end{align}
where $\xi_i$'s,  $\zeta_i$'s as well as $\phi_i$'s are all arbitrary phases.  Therefore, the PMNS matrix reads $U_\mathrm{PMNS}=E^\dagger N= Q D_c^\dagger N$. Note that $U$ and $D$ are related via 
\begin{align}
U^\dagger D &= \mathrm{diag}(e^{i \phi_1},e^{i \phi_2},e^{i \phi_3})V_\mathrm{CKM}\mathrm{diag}(e^{i \phi_4},e^{i \phi_5},1),
\label{eq:CKM}
\end{align}
where $V_\mathrm{CKM}$ is the Cabibbo-Kobayashi-Maskawa (CKM) matrix. 

We finally obtain $M_N$ that is compatible with suppression of all relevant proton decay signatures and viable charged fermion mass generation in the form of
\begin{align}
M_N=M_N^T&=  a_0
\bigg\{
Y_X  D_c M_D^\mathrm{diag} P^*   M_E^\mathrm{diag} Q D_c^\dagger
- Y_X D_c (M_D^\mathrm{diag})^2 D_c^\dagger
\nonumber\\& \hspace{30pt}
+ D_c^*Q^T M_E^\mathrm{diag} P^\dagger  M_D^\mathrm{diag} D_c^T Y_X^T
-  D_c^* (M_D^\mathrm{diag})^2 D_c^T Y_X^T
\bigg\}.
\label{eq:nu:24}
\end{align}

Clearly, since $Y_T$ can be expressed in terms of $M_E$ and $M_D$ due to a need to suppress proton decay signatures of the triplet, a numerical fit to the neutrino oscillation parameters allows one to determine the form of $Y_X$ matrix up to an overall scale factor.
We accordingly note that if one takes only the $d=4$ contribution towards $Y_X$ of Eq.\ \eqref{eq:Y_X}, that is skew-symmetric in the flavor space, the satisfactory numerical fit of neutrino parameters is not possible due to all additional constraints arising from the need to suppress partial proton decay lifetimes. However, if one also includes a $d=5$ term proportional to $Y_Z$, a satisfactory numerical solution does exist as we demonstrate next. Note that the numerical analysis is highly non-trivial since both $M_N$ and $U_\mathrm{PMNS}$ depend on the same unitary matrix $D_c$. 

We present, in what follows, a benchmark numerical fit, where input values for the neutrino sector are taken from Refs.~\cite{Esteban:2024eli,NUFIT}. We fit five observables, namely, the two neutrino mass-squared differences and the three mixing angles in the lepton sector. It is important to point out that our benchmark solution is only meant to serve as a proof of phenomenological viability of the extension under consideration. 

If we parametrize $D_c$ to be
\begin{align}
D_c=  \mathrm{diag}(e^{i\chi^{D_c}_1},e^{i\chi^{D_c}_2},e^{i\chi^{D_c}_3}) V(\theta^{D_c}_{ij},\delta^{D_c}) \mathrm{diag}(e^{i\alpha^{D_c}},e^{i\beta^{D_c}},1),  
\end{align}
where $V(\theta^{D_c}_{ij},\delta^{D_c})$ is a unitary matrix that depends on three angles and one phase as in the PDG convention for the CKM matrix and, furthermore, assume that the matrix elements of $Y_{X}$ are all real numbers, we obtain the following numerical fit:
\begin{align}
&a_0 Y_{X 11}= 2.67891\times 10^{-9} \;\mathrm{GeV}^{-1},
\\
&
Y_{X}= Y_{X 11}
\left(
\begin{array}{ccc}
 1. & -0.0164852 & -0.000162649 \\
 1.93483 & 1.32078 & -0.0000175168 \\
 0.960823 & 0.301348 & 0.0000197512 \\
\end{array}
\right),
\\&
(\xi_1,\xi_2,\xi_3)= (0.190099, 0.584088, 0.0202088),
\\&
(\zeta_1,\zeta_2,\zeta_3)= (2.96146, 1.37486, 1.9533),
\\&
(\theta^{D_c}_{12},\theta^{D_c}_{23},\theta^{D_c}_{13})= (0.0539512, 0.000434082, 0.000108572),
\\&
(\chi^{D_c}_1,\chi^{D_c}_2,\chi^{D_c}_3)= (-3.04041, 0.24154, 0.103072),
\\&
(\alpha^{D_c},\beta^{D_c},\delta^{D_c})= (0.930371, 0.452762, 0.214225).
\end{align}
Neutrino observables corresponding to this parameter set are summarized in the second column of Table~\ref{tab:neutrino_solutions}. Clearly, an excellent fit to the neutrino oscillation data, consisting of five observables, is obtained with a total $\chi^2=1.53$. This fit is close to the ruled out bound from cosmological data~\cite{ParticleDataGroup:2024cfk} that suggests $\sum m_i < 87$\,meV or $\sum m_i < 120$\,meV, depending on experiments included. Namely, our fit yields $\sum m_i = 76$\,meV. Moreover, neutrinoless double beta decay parameter, $m_{\beta\beta}=| \sum_i U^2_{ei}m_i |=2.69$\,meV, is also not too far from experimental bound of ($28$\textendash$122$)\,meV~\cite{KamLAND-Zen:2024eml}. 

\begin{table}[t]
\centering
\begin{longtable}{@{}lcc@{}}
\toprule
\textbf{Observables} &  $24_H$+$10_H$+$5_H$ &  $24_H$+$15_H$+$5_H$ \\
\midrule
\endfirsthead
\multicolumn{3}{c}{{\tablename\ \thetable{} -- continued from previous page}} \\
\toprule
\textbf{Observables} & \textbf{\boldmath $24_H + 10_H$} & \textbf{\boldmath $24_H + 15_H$} \\
\midrule
\endhead

\bottomrule
\endfoot

$\Delta m^2_{21}\times 10^{5}$ (eV$^2$) & 7.492 & 7.5085 \\
$\Delta m^2_{31}\times 10^{3}$ (eV$^2$) & 2.5349 & 2.5339 \\
$\sin^2\theta_{12}^{\mathrm{PMNS}}$ & 0.3075 & 0.3071 \\
$\sin^2\theta_{23}^{\mathrm{PMNS}}$ & 0.4653 & 0.4653 \\
$\sin^2\theta_{13}^{\mathrm{PMNS}}$ & 0.02191 & 0.02183 \\
\hline
$\chi^2$ & 1.53 & 1.57 \\ \hline
$m_1$ (eV) & 0.01075 & 0.00164 \\
$m_2$ (eV) & 0.01380 & 0.0088 \\
$m_3$ (eV) & 0.05148 & 0.0503 \\
$\delta_{\mathrm{CP}}^{\mathrm{PMNS}}$ (deg) & 176.12 & 128.51 \\
$\sum m_i$ (meV) & 76.0 & 60.7 \\
$m_{\beta\beta}$ (meV) & 2.69 & 1.01 \\
\hline

\end{longtable}
\caption{Benchmark fits of neutrino masses and mixing parameters for two different scenarios. Input values of neutrino observables, $\Delta m^2_{kl}$ and $\sin^2\theta_{ij}^{\mathrm{PMNS}}$, are taken from Refs.~\cite{Esteban:2024eli,NUFIT}. }
\label{tab:neutrino_solutions}
\end{table}

It is easy to understand why our numerical fit yields $Y_X$ that exhibits somewhat inverse hierarchy in the sense that $|Y_{X11}| \sim |Y_{X22}|  \gg |Y_{X33}|$. This happens due to the fact that $Y_X$ in Eq.\ \eqref{eq:nu:24} needs to compensate for highly hierarchical matrix $D_c (M_D^\mathrm{diag})^2 D_c^\dagger$, where the dominant entry is generated by the $(M_D^\mathrm{diag})^2_{33}$ element. 

We finally address the question of the cutoff scale $\Lambda$ in the $24_H$-$10_H$-$5_H$ scenario. Namely there is an upper bound on $\Lambda$ that originates from the fact that the up-type quark mass matrix $M_U$ needs to have substantial skew-symmetric component~\cite{Dorsner:2024seb} if one is to implement first condition presented in Eq.\ \eqref{eq:pdsuppression} with perturbative Yukawa couplings in Eqs.\ \eqref{eq:uncorrected_U_24} and \eqref{eq:qq_b_24}. The relevant analysis was described in detail in Ref.\ \cite{Dorsner:2024seb} and, for example, it yields $\Lambda < 57 M^\mathrm{max}_\mathrm{GUT}$ for the first unification scenario listed in  Table~\ref{tab:unification10H}.

\subsubsection{Extension with a 15-dimensional scalar representation}

\begin{figure}[th!]
\centering
\includegraphics[width=0.7\linewidth]{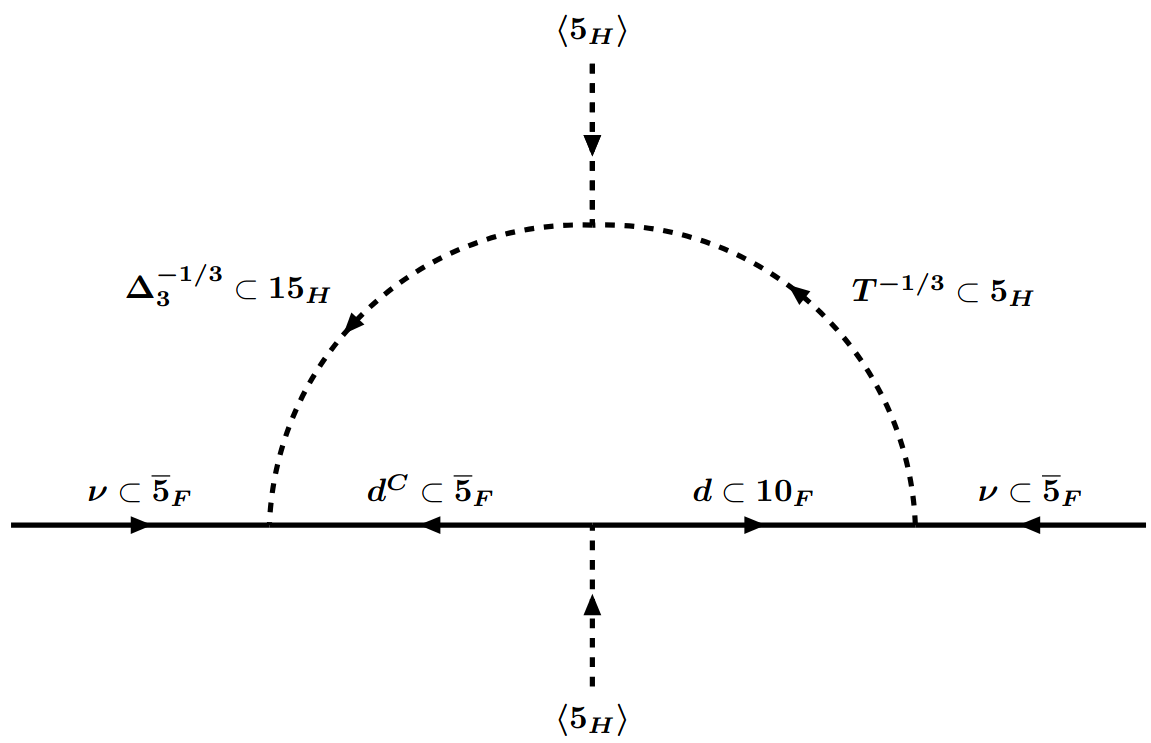}
\caption{One-loop neutrino mass generating diagram in the $15_H$ extension, when the $SU(5)$ symmetry breaking is accomplished with representations  $24_H$ and $5_H$.}
\label{fig:Nu15}
\end{figure}

If the Georgi-Glashow model is extended with a $15$-dimensional representation $15_H$, the neutrino mass generation can happen at the tree-level via the type-II seesaw mechanism~\cite{Minkowski:1977sc,Yanagida:1979as,Glashow:1979nm,Gell-Mann:1979vob,Mohapatra:1979ia,Schechter:1980gr,Schechter:1981cv}. (For tree-level neutrino mass generation in the context of $SU(5)$ framework, see, for example, Refs.~\cite{Dorsner:2005fq,Dorsner:2005ii,Bajc:2006ia,Dorsner:2006hw,Dorsner:2007fy,Antusch:2021yqe,Antusch:2022afk,Calibbi:2022wko,Antusch:2023kli,Antusch:2023mqe,Kaladharan:2024bop}.) Since the decomposition of $15_H$ reads
\begin{align}
 \Delta\equiv   &15_{H}=\Delta_{1}(1,3,1)+\Delta_{3}(3,2,1/6)+\Delta_{6}(6,1,-2/3),
\end{align}
one can note that this $SU(5)$ representation also has a color triplet leptoquark, i.e., $\Delta_{3}^{-1/3} \in \Delta_{3}(3,2,1/6)$, which can contribute towards neutrino masses at the one-loop level through the mixing with $T^{-1/3} \in 5_H$, as shown in Fig.~\ref{fig:Nu15}. 
The relevant mixing is provided by the following term in the scalar potential \begin{align}
    V\supset \mu\; 5^*_H 5^*_H 15_H \supset \mu v_5\; T^{1/3} \Delta^{-1/3}_3.
\end{align}

The mass-squared matrix for scalar leptoquarks $T^{-1/3} \in 5_H$ and $\Delta^{-1/3}_3 \in 15_H$ reads
\begin{align}
\label{eq:msquared15}
M^2_{S^\prime}=    \begin{pmatrix}
m^2_T&\mu v_5
\\
\mu v_5&m^2_{\Delta_3}
\end{pmatrix},
\end{align}
where the mixing angle between $T^{-1/3}$ and $\Delta^{-1/3}_3$ is
\begin{align}
    \tan 2\theta^\prime=\frac{ 2\mu v_5 }{m^2_T-m^2_{\Delta_3}}.
\end{align}

We explicitly assume that the one-loop contribution of Fig.~\ref{fig:Nu15} dominates over the tree-level contribution in what follows. (Note that $T^{-1/3} \in 5_H$ along with $\Delta_6(6,1,-2/3) \in 15_H$ can provide neutrino mass of the two-loop order via the Zee-Babu diagram~\cite{Zee:1980ai,Cheng:1980qt,Babu:1988ki}. However, one-loop diagram dominates over the aforementioned two-loop contribution.) 

With the introduction of  $15_H$, we have additional Yukawa couplings that play role in neutrino mass generation
\begin{align}
   -\mathcal{L}_Y \supset Y^{\alpha \beta}_{Y^\prime} \overline 5_{F i}^\alpha \overline 5_{F j}^\beta 15_H^{ij}  +\frac{1}{\Lambda}  Y^{\alpha \beta}_{Z^\prime} \overline 5_{F i}^\alpha \overline 5_{F j}^\beta 15_H^{ik} 24^{j}_{H k}
    \supset   \nu^TC^{-1} Y_{X^\prime}  d^c_0 \Delta_3^{-1/3},
\end{align}
where we include the $d=5$ contraction and define the following effective coupling matrix: 
\begin{align}
 Y_{X^\prime}=  -\sqrt{2}Y_{Y^\prime} -   \frac{\epsilon_{24}}{2\sqrt{2}} 
Y_{Z^\prime}.
\end{align}
Here, $Y_{Y^\prime}$ is a symmetric matrix in the flavor space, whereas $Y_{Z^\prime}$ is an arbitrary matrix.

Even though the gauge coupling unification, within this particular scenario, does not require presence of higher-dimensional contributions towards kinetic terms for the gauge fields~\cite{Dorsner:2005fq}, their inclusion somewhat helps~\cite{Senjanovic:2024uzn}, especially in the light triplet regime, to increase upper limit on $M_\mathrm{GUT}$. We accordingly provide  in Table~\ref{tab:unification15H} the highest possible unification scale $M^\mathrm{max}_\mathrm{GUT}$ as a function of degenerate masses $m_{S_1}=m_{S_2}$ of linear combinations of scalar leptoquarks $T^{-1/3}$ and $\Delta_3^{-1/3}$ that correspond to physical states $S_1$ and $S_2$ and dimensionless parameter $\epsilon_5$ of Eq.\ \eqref{eq:epsilon_5}.  
\begin{table} 
\centering
\begin{tabular}{|c|c|c|c|}
  \hline
$\epsilon_5$ & $m_{S_1}=m_{S_2}$ (TeV) & $M^\mathrm{max}_\mathrm{GUT}$  (10$^{14}$\,GeV) & $\alpha_\mathrm{GUT}^{-1}$ \\   
\hline \hline 
$0.020$ & $10^0$ & $5.957$ & $38.1$ \\ 
$0.021$  & $10^1$ & $5.322$ & $38.2$ \\
$0.021$  & $10^2$ & $4.786$ & $38.3$ \\ 
$0.021$ & $10^3$ & $4.245$ & $38.4$ \\ 
$0.022$  & $10^4$ & $3.753$ & $38.5$ \\  
$0.022$  & $10^5$ & $3.375$ & $38.6$ \\  
  \hline
\end{tabular}
\caption{The highest possible unification scale $M^\mathrm{max}_\mathrm{GUT}$ as a function of degenerate masses of linear combinations of scalars $T^{-1/3}$ and $\Delta_3^{-1/3}$ within the $15_H$ extension of the $24_H$ scenario.}
\label{tab:unification15H}
\end{table}

Since we have viable gauge coupling unification that requires suppression of both the scalar and gauge boson mediated proton decay signatures, we assume that the same conditions we imposed on the $10_H$ extension are also at play here in order to have phenomenologically viable scenario. These conditions are specified in Eqs.\ \eqref{eq:spdsuppression} and \eqref{eq:pdsuppression} for scalar and gauge boson mediation, respectively. This also means that an upper bound on $\Lambda$, once again, originates from the fact that the up-type quark mass matrix $M_U$ needs to have substantial skew-symmetric component~\cite{Dorsner:2024seb}. We accordingly find, for the first unification scenario listed in  Table~\ref{tab:unification15H} that $\Lambda < 57 M^\mathrm{max}_\mathrm{GUT}$.

We finally present an example benchmark fit, where we consider a scenario when the $d=4$ term towards $Y_{X^\prime}$ dominates. More specifically, we consider a scenario when $Y_{X^\prime}$ is a symmetric matrix with real elements. Since the relevant neutrino mass matrix reads
\begin{align}
M_N=M_N^T&=  a^\prime_0
\bigg\{
Y_{X^\prime}  D_c M_D^\mathrm{diag} P^*   M_E^\mathrm{diag} Q D_c^\dagger
- Y_{X^\prime} D_c (M_D^\mathrm{diag})^2 D_c^\dagger
\nonumber\\& \hspace{30pt}
+ D_c^*Q^T M_E^\mathrm{diag} P^\dagger  M_D^\mathrm{diag} D_c^T Y_{X^\prime}^T
-  D_c^* (M_D^\mathrm{diag})^2 D_c^T Y_{X^\prime}^T
\bigg\}
\label{eq:nu:2415},
\end{align}
where 
\begin{align}
    a^\prime_0=\frac{3\sqrt{2}\sin2\theta^\prime}{16\pi^2 v_5} \ln\left( \frac{m_{S_1}}{m_{S_2}} \right),
\end{align}
our numerical fit yields
\begin{align}
&a^\prime_0 Y_{X^\prime 11}= 3.58225\times 10^{-9}  \;\mathrm{GeV}^{-1},
\\
&
Y_{X^\prime}= Y^\prime_{X 11}
\left(
\begin{array}{ccc}
 1. & 1.21286 & 0.00032875 \\
 1.21286 & 0.581634 & -0.000276709 \\
 0.00032875 & -0.000276709 & -0.000018021 \\
\end{array}
\right),
\\&
(\xi_1,\xi_2,\xi_3)= (0.986652, 1.14213, 0.0772671),
\\&
(\zeta_1,\zeta_2,\zeta_3)= (2.60532, 1.98709, 0.597686),
\\&
(\theta^{D_c}_{12},\theta^{D_c}_{23},\theta^{D_c}_{13})= (0.118582, 0.00066669, 0.000432764),
\\&
(\chi^{D_c}_1,\chi^{D_c}_2,\chi^{D_c}_3)= (-3.02107, 2.37879, -0.830596),
\\&
(\alpha^{D_c},\beta^{D_c},\delta^{D_c})= (1.1893, 0.379583, -2.62208).
\end{align}
Neutrino observables corresponding to this numerical fit are summarized in the third column of Table~\ref{tab:neutrino_solutions}. One can observe that, once again, $|Y_{X^\prime 11}| \sim |Y_{X^\prime 22}|  \gg |Y_{X^\prime 33}|$, in agreement with our discussion of the numerical fit within the $10_H$ extension.

Before we conclude this section we briefly comment on potentially problematic proton decay signatures that might be induced by the mixing between leptoquark multiplets in either $10_H$ or $15_H$ with the leptoquark in $5_H$, since this mixing is essential for the generation of viable neutrino masses and thus must be present. These proton decay signatures, however, do not exist in both extensions under consideration since we insist on the suppression of the quark-quark interactions of leptoquark $T^{-1/3} \in 5_H$. This means that leptoquark multiplets $\eta_3 \in 10_H$ and $\Delta_3 \in 15_5$ as well as leptoquark $T \in 5_H$ exclusively couple to the quark-lepton pairs. The only contribution towards proton decay might come from the triple-leptoquark interaction~\cite{Dorsner:2022twk,Dorsner:2024gmy} between $\eta_3 \in 10_H$ and $T \in 5_H$ via the $10_H$-$10_H$-$5_H$ contraction, but  that particular interaction is not needed for the fermion mass generation at all.

\subsection{The $75_H$ scenario case studies}

\subsubsection{Extension with a 10-dimensional scalar representation}

\begin{figure}[th!]
\centering
\includegraphics[width=0.7\linewidth]{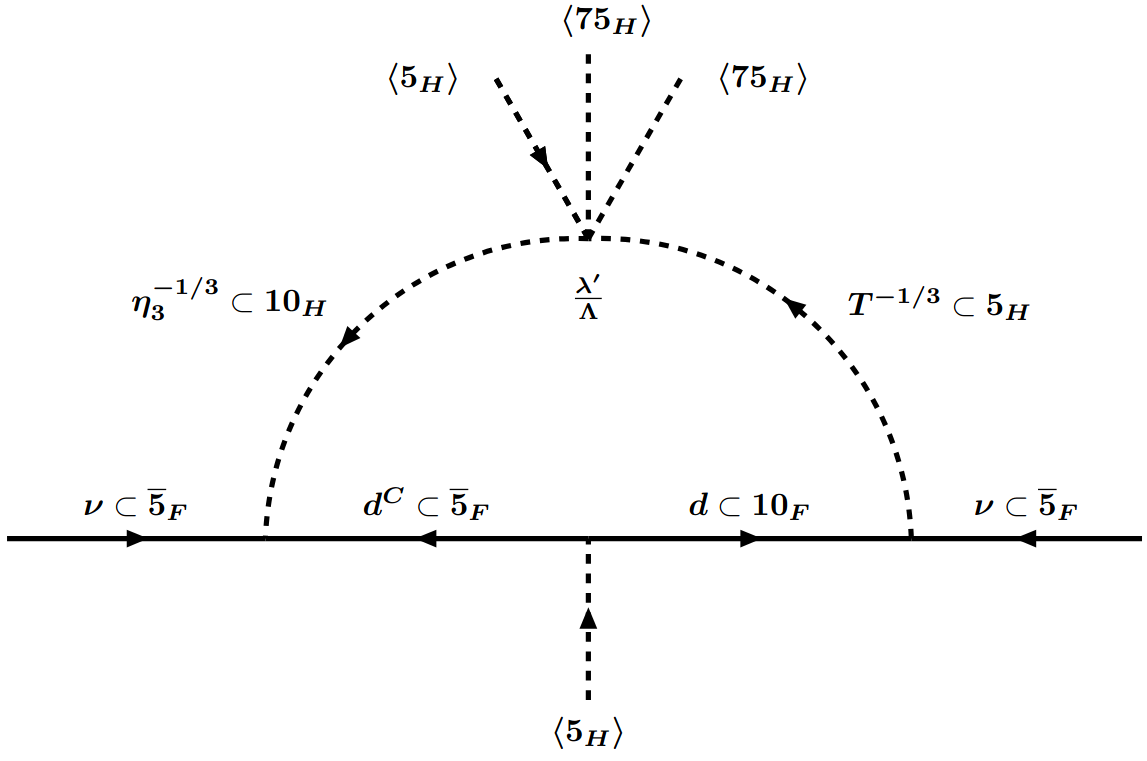}
\caption{One-loop neutrino mass generating diagram in the $10_H$ extension, when the $SU(5)$ symmetry breaking is accomplished with representations  $75_H$ and $5_H$.}
\label{fig:Nu1075}
\end{figure}

First, we point out one crucial difference between the $24_H$+$10_H$+$5_H$ and $75_H$+$10_H$+$5_H$ scenarios. Namely, in the former scenario, the mixing between the scalar leptoquarks that is needed to provide non-zero neutrino masses appears at the $d=4$ level. The corresponding scalar mixing for the latter scenario actually first appears at the $d=5$ level, as can be seen in Fig.~\ref{fig:Nu1075}. It is thus crucial to go beyond the $d=4$ contractions if one is to explain neutrino masses and mixing parameters. The relevant $d=5$ term in the scalar potential takes the following form:
\begin{align}
    V \supset  \frac{\lambda^\prime }{\Lambda} \; 5^*_{H i} 5^*_{H j} 10_H^{ik} 75^{mn}_{H kl} 75^{lj}_{H mn} \supset -\frac{4}{9}\lambda^\prime v_5 v_{75} \epsilon_{75} T^{1/3} \eta^{-1/3}_3.
\end{align} 
We can now introduce the mass-squared matrix for the scalar leptoquarks via 
\begin{align}
\label{eq:msquared1075}
M^2_S=    \begin{pmatrix}
m^2_T&-\frac{4}{9}\lambda^\prime v_5 v_{75} \epsilon_{75}
\\
-\frac{4}{9}\lambda^\prime v_5 v_{75} \epsilon_{75}&m^2_{\eta_3}
\end{pmatrix},
\end{align}
where the mixing angle reads 
\begin{align}
    \tan 2\theta^{\prime \prime}=\frac{ -8\lambda^\prime v_5 v_{75} \epsilon_{75}/9 }{m^2_T-m^2_{\eta_3}}.
\end{align}
Here, again, we can use $m_{S_1}^2$ and $m_{S_2}^2$ to denote squares of the masses of physical eigenstates of the leptoquarks after the mixing between $T^{-1/3}$ and $\eta_3^{-1/3}$.

The left vertex of Fig.~\ref{fig:Nu1075} is generated via
\begin{align}
   -\mathcal{L}_Y& \supset Y_{Y \alpha \beta} \overline 5_{F i}^\alpha \overline 5_{F j}^\beta 10_H^{ij}  +\frac{1}{\Lambda}  Y_{Z \alpha \beta} \overline 5_{F i}^\alpha \overline 5_{F j}^\beta 10_H^{kl} 75^{ij}_{H kl}
   +\frac{1}{\Lambda^2}  Y_{W_1 \alpha \beta} \overline 5_{F i}^\alpha \overline 5_{F j}^\beta 10_H^{kl} 75^{ij}_{H mn}75^{mn}_{H kl}
   \nonumber\\&
   +\frac{1}{\Lambda^2}  Y_{W_2 \alpha \beta} \overline 5_{F i}^\alpha \overline 5_{F j}^\beta 10_H^{kl} 75^{im}_{H kn}75^{jn}_{H lm}
   +\frac{1}{\Lambda^2}  Y_{W_3 \alpha \beta} \overline 5_{F i}^\alpha \overline 5_{F j}^\beta 10_H^{ik} 75^{jl}_{H mn}75^{mn}_{H kl},
\end{align}
where we included all $d=4$, $d=5$, and $d=6$ contractions.

The relevant  interaction between $\eta^{-1/3} \in 10_H$ and the $d^C$-$\nu$ pairs, in the flavor basis, reads
\begin{align}
 Y_X&=  \sqrt{2}Y_{Y} -   \frac{2\epsilon_{75}}{3} 
Y_{Z}
+   \frac{2\sqrt{2}\epsilon^2_{75}}{9} 
Y_{W_1}
-   \frac{\sqrt{2}\epsilon^2_{75}}{3} 
Y_{W_2}
-\frac{4\sqrt{2}\epsilon^2_{75}}{9} 
Y_{W_3},
\end{align}
where $Y_Y$, $Y_Z$, $Y_{W_1}$, and $Y_{W_2}$ are skew-symmetric matrices in the flavor space, whereas $Y_{W_3}$ is an arbitrary matrix. With this, the neutrino mass matrix is determined by Eq.~\eqref{eq:numass}, where one should replace $\theta$ with $\theta^{\prime \prime}$ and insert $Y_T$ as given in the square brackets of Eq.\ \eqref{scalar-3-75}, after one imposes conditions of Eqs.\ \eqref{eq:qq_a_75} and \eqref{eq:qq_b_75} to have a light triplet. In fact, $Y_T$ is especially simple in the $75_H$ scenario with a light triplet as it reads
\begin{align}
Y_T=\frac{\sqrt{2}}{v_5}M_D^T.  \label{eq:triplet75}
\end{align}

We note that the gauge coupling unification, at sufficiently large $M_\mathrm{GUT}$, can be trivially achieved within this particular  extension due to the fact that the scalar multiplet $\Phi_3 (8,3,0) \in 75_H$ is extremely beneficial in that regard. For example, the highest possible value of $M_\mathrm{GUT}$ for $m_{S_1}=m_{S_2}=1\,\mathrm{TeV}$ is $M^\mathrm{max}_\mathrm{GUT}=1.0 \times 10^{19}$\,GeV. As in all other scenarios, our automated unification procedure looks for the highest possible unification scale $M^\mathrm{max}_\mathrm{GUT}$ by treating the masses of all other scalars to be free parameters that can take any value between $1$\,TeV and $M_\mathrm{GUT}$. It is thus clear that there is no need to suppress gauge boson mediated proton decay at all. This, in turn, enables one to trivially accommodate observed masses of all the SM fermions. More specifically, since the unitary matrix $E$ is not restricted in any way, it can always be redefined via $E=N U_\mathrm{PMNS}^\dagger$, where $N$ takes $M_N$, given by  
\begin{align}
\label{eq:numass75}
M_N  \approx \frac{3\sin2\theta^{\prime\prime}}{32\pi^2} \ln\left( \frac{m_{S_1}^2}{m_{S_2}^2} \right) \bigg\{
Y_X D_c \left(M_D^\mathrm{diag}\right)^2 D_c^\dagger  + D^*_c  \left(M_D^\mathrm{diag}\right)^2 D^T_c Y_X^T
\bigg\}
\end{align}
into a diagonal form. Consequently, all one needs to do in order to prove viability of this extension is to fit the two mass-squared differences in the neutrino sector, 
which can be trivially accomplished even with a skew-symmetric matrix $Y_X$. 

The upper bound on a cutoff $\Lambda$, in the $75_H$-$10_H$-$5_H$ scenario, originates from a need to simultaneously generate phenomenologically viable $M_E$ and $M_D$ while imposing condition given in Eq.\ \eqref{eq:qq_a_75} in order to suppress quark-quark-leptoquark couplings. This should be contrasted with the $24_H$-$10_H$-$5_H$ scenario, where $\Lambda$ is bounded from above due to a need to have substantial skew-symmetric form of $M_U$. To produce a conservative estimate of an upper bound on $\Lambda$ for the $75_H$-$10_H$-$5_H$ scenario, we set, for simplicity, matrix $Y_d$ in Eq.\ \eqref{eq:lagrangian_75} to zero to obtain, using Eqs.\ \eqref{eq:uncorrected_E_75}, \eqref{eq:uncorrected_D_75}, and \eqref{eq:qq_a_75}, the following identity 
\begin{align}
    \frac{1}{6} Y_c \epsilon^2_{75} v_5 = -\frac{3}{8}(M^T_D+M_E).
    \label{eq:cutoff_75H}
\end{align}
It is this identity that yields an upper bound on $\Lambda$ if one demands that the entries in matrix $|Y_c/6|$, as featured in Eqs.\ \eqref{eq:uncorrected_E_75} and \eqref{eq:uncorrected_D_75}, do not exceed perturbativity limit that we take to be $\sqrt{4 \pi}$. For example, if one conservatively takes the dominant entry of the right-hand side of Eq.\ \eqref{eq:cutoff_75H} to be $3/8 (m_b+m_\tau)$, one obtains that $\Lambda < 30 v_{75}$. Again, this is very conservative estimate that, nevertheless, demonstrates self-consistancy of the $75_H$-$10_H$-$5_H$ scenario. Since $M_\mathrm{GUT}$ can be very large, $\Lambda$ can even be identified with the Planck scale.

\subsubsection{Extension with a 15-dimensional scalar representation}

The neutrino mass diagram of interest is practically the same as the one already shown in  Fig.~\ref{fig:Nu15}. Its left vertex is generated through the following $d=4$ and $d=6$ contractions 
\begin{align}
   -\mathcal{L}_Y& \supset Y_{Y^\prime \alpha \beta} \overline 5_{F i}^\alpha \overline 5_{F j}^\beta 15_H^{ij}  
   +\frac{1}{\Lambda^2}  Y_{W_1^\prime \alpha \beta} \overline 5_{F i}^\alpha \overline 5_{F j}^\beta 15_H^{kl} 75^{im}_{H kn}75^{jn}_{H lm}
   +\frac{1}{\Lambda^2}  Y_{W_2^\prime \alpha \beta} \overline 5_{F i}^\alpha \overline 5_{F j}^\beta 15_H^{ik} 75^{jl}_{H mn}75^{mn}_{H kl},
\end{align}
where the effective coupling of the triplet $\Delta^{-1/3} \in 15_H$ with the $d^C$-$\nu$ pairs, in the flavor basis, is
\begin{align}
 Y_{X^\prime}&=  -\sqrt{2}Y_{Y^\prime} +   \frac{2\sqrt{2}\epsilon^2_{75}}{9} 
Y_{W_1^\prime}
+\frac{-4\sqrt{2}\epsilon^2_{75}}{9} Y_{W_2^\prime}.
\end{align}
Here, $Y_{Y^\prime}$ and $Y_{W_1^\prime}$ are symmetric matrices, whereas $Y_{W_2^\prime}$ is an arbitrary matrix. The neutrino mass matrix is determined by Eq.~\eqref{eq:numass}, where one would need to insert $Y_{X^\prime}$ instead of $Y_X$ and use Eq.\ \eqref{eq:triplet75} for $Y_T$.

Since the gauge coupling unification can happen at sufficiently large $M_\mathrm{GUT}$ that does not require any suppression of the gauge boson mediated proton decay, one can, similarly to the $75_H$+$10_H$+$5_H$ scenario, trivially accommodate fermion masses and mixing parameters within the light triplet regime. Also, an upper bound on the cutoff $\Lambda$ for the $75_H$+$15_H$+$5_H$ scenario has exactly the same origin as for the $75_H$+$10_H$+$5_H$ scenario.

We summarize our findings as follows. The $24_H$+$10_H$+$5_H$ scenario requires corrections to the gauge kinetic terms in order to provide gauge coupling unification, where its viability also needs suppression of gauge mediated proton decay. The $24_H$+$15_H$+$5_H$ scenario can unify without the gauge kinetic term corrections but still needs suppression of the gauge mediated proton decay signatures. The $75_H$+$10_H$+$5_H$ and $75_H$+$15_H$+$5_H$ scenarios, on the other hand, can both yield high enough unification scale that does not require any suppression of gauge mediated proton decay. The upper bound on the cutoff $\Lambda$, for the $24_H$+$10_H$+$5_H$ and $24_H$+$15_H$+$5_H$ scenarios, originates from a need to suppress gauge mediated proton decay signatures and is primarily determined by the features of the up-type quark sector. The upper bound on $\Lambda$ for the $75_H$+$10_H$+$5_H$ and $75_H$+$15_H$+$5_H$ scenarios, on the other hand, is directly related to a necessary use of $d=6$ operators in the charged lepton and down-type quark sectors.  

\section{Experimental implications}
\label{sec:colliders}

To showcase the experimental potential of the light color triplet regime, we concentrate on the signatures of the most constraining scenario comprising $24_H$, $10_H$, and $5_H$. 

There are three leptoquarks in the $24_H$+$10_H$+$5_H$ scenario. These are $\eta_3^{2/3} \in \eta_3 (3,2,1/6) \in 10_H$, $\eta_3^{-1/3} \in \eta_3 (3,2,1/6) \in 10_H$, and $T^{-1/3}(3,1,-1/3) \in 5_H$, where $\eta_3^{-1/3}$ and $T^{-1/3}$ need to mix, as given in Eq.\ \eqref{eq:linear_comb}, in order to generate neutrino masses at the one-loop level. 

The scalar leptoquark interactions of $\eta_3 (3,2,1/6)$ and $T^{-1/3}$, in the $24_H$+$10_H$+$5_H$ scenario, are
\begin{align}
-\mathcal{L}_Y &\supset 
u^T_\alpha C^{-1}e_\beta T^{1/3}  \bigg\{ -U^T Y_T E \bigg\}_{\alpha\beta}
+
d^T_\alpha C^{-1}\nu_\beta T^{1/3}  \bigg\{ D^T Y_T N
\bigg\}_{\alpha\beta} 
\nonumber\\
&+
\nu^T_\alpha C^{-1}d^c_\beta \eta_3^{-1/3}  \bigg\{ N^T Y_X D_c \bigg\}_{\alpha\beta}
+
e^T_\alpha C^{-1}d^c_\beta \eta_3^{2/3}  \bigg\{- E^T Y_X D_c \bigg\}_{\alpha\beta}\nonumber\\
&+u^{C,T}_{\alpha}C^{-1}e^C_\beta T^{-1/3} \bigg\{U^\dagger_c \bigg[
 \frac{5}{2}\left(Y^T_4-Y_4 \right) \epsilon_{24}\bigg]
 E_c^*
\bigg\}_{\alpha\beta},
\label{eq:LHC}
\end{align}
where $Y_X$ and $Y_T$ are given in Eqs.\ \eqref{eq:Y_X} and \eqref{eq:Y_T10_H}, respectively.

Since $N$, $Q$, $D_c$, and, consequentially, $E\equiv D_c Q^\dagger$ are all determined from the neutrino fit, we can reconstruct Yukawa couplings of $\eta_3^{-1/3}$ and $\eta_3^{2/3}$, up to an overall scale. Namely, from the benchmark fit provided in Sec.\ \ref{sec:neutrinos}, we have
\begin{align}
&|N^T Y_X D_c|= |Y_{X 11}| \left(
\begin{array}{ccc}
 0.888 & 0.591 & 0.00028 \\
 0.350 & 0.680 & 0.00033 \\
 2.245 & 0.855 & 0.00015 \\
\end{array}
\right),
\\
&|E^T Y_X D_c|=|Y_{X 11}| \left(
\begin{array}{ccc}
 0.976 & 0.250 & 0.000062 \\
 1.946 & 1.217 & 0.00037 \\
 1.103 & 0.0181 & 0.000259 \\
\end{array}
\right),
\end{align}
where we clearly see that both components of $\eta_3$ couple most strongly to the $d$ quark. Also, the form of $|E^T Y_X D_c|$ stiplulates that $\eta_3^{2/3}$ would preferentially decay into muons and light jets, if produced at colliders.

The situation with $\eta_3^{-1/3}$ and $T^{-1/3}$ is more involved, even if one neglects the effect of their mixing via angle $\theta$ of Eq.\ \eqref{eq:linear_comb}. First thing to note is that the interactions of $T^{-1/3}$ that are proportional to $Y_T$, i.e., the couplings in the first line of Eq.\ \eqref{eq:LHC}, are completely irrelevant for our discussion as they are proportional to the Yukawa couplings of the down-type quarks and charged leptons. Again, this makes them completely negligible for the discussion of the $T^{-1/3}$ production mechanisms and/or decay signatures. What is relevant, though, is the interactions of $T^{-1/3}$ with the up-type quarks and charged leptons in the last line of Eq.\ \eqref{eq:LHC}. Namely, these couplings cannot all be small since $Y_4$ has to exhibit substantial skew-symmetric properties in order for the first condition in Eq.\ \eqref{eq:pdsuppression} to hold. Note that the symmetric form of $M_U$ in Eq.\ \eqref{eq:uncorrected_U_24} would imply that $U_c$ and $U$ are one and the same matrix, which would be in conflict with Eq.\ \eqref{eq:pdsuppression}. In fact, it is the skew-symmetricity of $Y_4$, in combination with the need for perturbativity, that places the most stringent bound on the cutoff scale $\Lambda$, as discussed in detail in Ref.\ \cite{Dorsner:2024seb}. What one can thus say with certainty is that $T^{-1/3}$ will couple strongly to the up-type quarks and charged leptons, whereas $\eta_3^{-1/3}$ will preferential couple to the $d$ quark and a neutrino, where the overall scale, given by $|Y_{X11}|$, is not known.

It is not guaranteed, even in the $24_H$+$10_H$+$5_H$ scenario, that the three leptoquarks in question will be accelerator accessible. We can note, however, that the unification scale $M^\mathrm{max}_\mathrm{GUT}$ is increased as the masses of leptoquarks are lowered. This simply mean that any future improvement in proton decay lifetime limits will improve upper limit on the leptoquark masses within both the $24_H$+$10_H$+$5_H$ and $24_H$+$15_H$+$5_H$ scenarios. 

\section{Conclusions}
\label{sec:conclusions}

We present a novel perspective on a long-standing issue of the doublet–triplet splitting problem within the $SU(5)$ framework. Our proposal allows for a color scalar of the doublet–triplet splitting notoriety to be light without any conflict with experimental bounds on partial proton decay lifetimes. We explicitly demonstrate, through introduction of higher-dimensional operators, how to suppress dangerous baryon number violating couplings associated with the color triplet mediation if the $SU(5)$ gauge group is broken down to $SU(3) \times SU(2) \times U(1)$ by either a $24$-dimensional or a $75$-dimensional representation while $SU(3) \times SU(2) \times U(1)$ is subsequently broken down to $SU(3) \times U(1)_\mathrm{em}$ by a $5$-dimensional representation. We compare the main features of these two distinct symmetry breaking scenarios and, for each of them, we further study two phenomenologically different paths towards viable neutrino mass generation, where the proposed one-loop level neutrino mass generation mechanism is tied to the lightness of the aforementioned color triplet scalar. One path requires introduction of an additional $10$-dimensional scalar representation, whereas the other one uses a single $15$-dimensional scalar representation.  This work highlights main features of a novel approach to the $SU(5)$ model building through consistent use of non-renormalizable operators, where light leptoquarks are not a liability but a powerful probe of new physics.

\section*{Acknowledgments}

I.D.\ would like to thank Damir Be\v cirevi\' c and Olcyr Sumensari for hospitality at IJCLab, Orsay, where part of this work has been performed. I.D.\ also acknowledges the financial support from the Slovenian Research Agency (research core funding No.\ P1-0035). S.S.\ would like to thank  Borut Bajc for discussion. S.S.\  acknowledges the financial support
from the Slovenian Research Agency (research core funding No. P1-0035 and N1-0321). 

\bibliographystyle{style}
\bibliography{references}
\end{document}